\documentclass[twocolumn]{aastex62}

\usepackage{amsmath}
\usepackage{mathtools}
\usepackage{enumitem}
\usepackage{graphicx}

\begin{document}

\title{Magnetic Fields and Afterglows of BdHNe: Inferences from GRB 130427A, GRB 160509A, GRB 160625B, GRB 180728A and GRB 190114C}

\author{J.~A.~Rueda}
\affiliation{ICRA, Dipartimento di Fisica, Sapienza Universit\`a di Roma, P.le Aldo Moro 5, 00185 Rome, Italy.}
\affiliation{ICRANet, P.zza della Repubblica 10, 65122 Pescara, Italy. yu.wang@icranet.org}
\affiliation{ICRANet-Rio, Centro Brasileiro de Pesquisas F\'isicas, Rua Dr. Xavier Sigaud 150, 22290--180 Rio de Janeiro, Brazil.}
\affiliation{ICRANet-Ferrara, Dipartimento di Fisica e Scienze della Terra, Universit\`a degli Studi di Ferrara, Via Saragat 1, I--44122 Ferrara, Italy}
\affiliation{Dipartimento di Fisica e Scienze della Terra, Universit\`a degli Studi di Ferrara, Via Saragat 1, I--44122 Ferrara, Italy}
\affiliation{INAF, Istituto di Astrofisica e Planetologia Spaziali, Via Fosso del Cavaliere 100, 00133 Rome, Italy.}

\author{Remo~Ruffini}
\affiliation{ICRA, Dipartimento di Fisica, Sapienza Universit\`a di Roma, P.le Aldo Moro 5, 00185 Rome, Italy.}
\affiliation{ICRANet, P.zza della Repubblica 10, 65122 Pescara, Italy. yu.wang@icranet.org}
\affiliation{ICRANet-Rio, Centro Brasileiro de Pesquisas F\'isicas, Rua Dr. Xavier Sigaud 150, 22290--180 Rio de Janeiro, Brazil.}
\affiliation{Universit\'e de Nice Sophia Antipolis, CEDEX 2, Grand Ch\^{a}teau Parc Valrose, Nice, France.}
\affiliation{INAF, Viale del Parco Mellini 84, 00136 Rome, Italy.}

\author{Mile~Karlica}
\affiliation{ICRA, Dipartimento di Fisica, Sapienza Universit\`a di Roma, P.le Aldo Moro 5, 00185 Rome, Italy.}
\affiliation{ICRANet, P.zza della Repubblica 10, 65122 Pescara, Italy. yu.wang@icranet.org}
\affiliation{Universit\'e de Nice Sophia Antipolis, CEDEX 2, Grand Ch\^{a}teau Parc Valrose, Nice, France.}

\author{Rahim~Moradi}
\affiliation{ICRA, Dipartimento di Fisica, Sapienza Universit\`a di Roma, P.le Aldo Moro 5, 00185 Rome, Italy.}
\affiliation{ICRANet, P.zza della Repubblica 10, 65122 Pescara, Italy. yu.wang@icranet.org}
\affiliation{INAF -- Osservatorio Astronomico d'Abruzzo,Via M. Maggini snc, I-64100, Teramo, Italy. rahim.moradi@inaf.it}

\author{Yu~Wang}
\affiliation{ICRA, Dipartimento di Fisica, Sapienza Universit\`a di Roma, P.le Aldo Moro 5, 00185 Rome, Italy.}
\affiliation{ICRANet, P.zza della Repubblica 10, 65122 Pescara, Italy. yu.wang@icranet.org}
\affiliation{INAF -- Osservatorio Astronomico d'Abruzzo,Via M. Maggini snc, I-64100, Teramo, Italy. rahim.moradi@inaf.it}


\begin{abstract}
GRB 190114C is the first binary-driven hypernova (BdHN) fully observed from the initial supernova appearance to the final emergence of the optical SN signal. It offers an unprecedented testing ground for the BdHN theory and it is here determined and further extended to additional gamma-ray bursts (GRBs). BdHNe comprise two subclasses of long GRBs with progenitors a binary system composed of a carbon-oxygen star (CO$_\textrm{core}$) and a neutron star (NS) companion. The CO$_\textrm{core}$ explodes as a SN leaving at its center a newborn NS ($\nu$NS). The SN ejecta hypercritically accretes both on the $\nu$NS and the NS companion. BdHNe I are the tightest binaries where the accretion leads the companion NS to gravitational collapse into a black hole (BH). In BdHN II the accretion onto the NS is lower, so there is no BH formation. We observe the same structure of the afterglow for GRB 190114C and other selected examples of BdHNe I (GRB 130427A, GRB 160509A, GRB 160625B) and for BdHN II (GRB 180728A). In all the cases the explanation of the afterglow is reached via the synchrotron emission powered by the $\nu$NS: their magnetic fields structures and their spin are determined. For BdHNe I, we discuss the properties of the magnetic field embedding the newborn BH, inherited from the collapsed NS and amplified during the gravitational collapse process, and surrounded by the SN ejecta.

\end{abstract}

\keywords{gamma-ray bursts: general --- binaries: general --- stars: neutron --- supernovae: general --- black hole physics}


\section{Introduction}\label{sec:1}
We first shortly review the traditional afterglow models and the possible alternatives. This task has been facilitated by the appearance of the comprehensive book by \citet{2018pgrb.book.....Z}. We focus on the additional results introduced since by the understanding: of the X-ray flare~\citep{2018ApJ...852...53R}, of the afterglow of GRB 130427A~\citep{2018ApJ...869..101R}, and  of GRB 190114C~\citep{2019ApJ...883..191R,2019arXiv190404162R}.

We first recall the well known discoveries by the Beppo-SAX satellite:
\begin{enumerate}[label=\alph*)]
\item 
the discovery of the first afterglow in GRB 970228, \citealp{1997Natur.387..783C});
\item 
the consequent identification of the cosmological redshift of GRBs (GRB 970508, \citealp{1997Natur.387..878M}) which proved the cosmological nature of GRBs and their outstanding energetics;
\item 
the first clear coincidence of a long GRB with the onset of a supernova (GRB 980425/SN 1998bw, \citealp{1998Natur.395..670G}).
\end{enumerate}

Even before these discoveries, three contributions, based on first principles, formulated models for long GRBs assuming their cosmological nature and originating from a BH formation. At the time, these works expressed the point of view of a small minority. A parallel successful move was done by Paczynski and collaborators for short GRBs \citep{1991AcA....41..257P,1992grbo.book...67P,1992ApJ...395L..83N}. The aforementioned three contributions are the following: 
\begin{enumerate}[label=\alph*)]
    \item 
    \citet{1975PhRvL..35..463D} predicted that vacuum polarization process occurring around an overcritical Kerr-Newman black hole (BH) leads toward GRB energetics of up to $10^{54}$~erg, linking their activities as well to the emergence of ultra-high energy cosmic rays (UHECRs);
    \item 
    the works of \citet{1992MNRAS.258P..41R,1538-4357-482-1-L29} also proposed a BH as the origin of GRBs but there, an ultra-relativistic blastwave, whose expansion follows the Blandford-McKee self-similar solution, was used to explain the prompt emission phase \citep{1976PhFl...19.1130B};
    \item 
    the work of \citet{1993ApJ...405..273W} linked the GRB origin to a Kerr BH emitting an ultra-relativistic jet originating from the accretion of toroidal material onto the BH. There, it was presented the idea that for long GRBs the BH would be likely produced from the direct collapse of a massive star, a ``failed'' SN leading to a large BH of approximately $5 M_{\odot}$, possibly as high as $10 M_{\odot}$,  a ``collapsar''.
\end{enumerate}

\subsection{Traditional afterglow model originating from  BH}

The paper by \citet{1975PhRvL..35..463D} has started only recently to attract attention in binary-driven hypernovae (BdHNe) in the context of the exact solution of the Einstein-Maxwell equations by \citet{1974PhRvD..10.1680W}, see section~\ref{sec:2} for further details. The papers by \citet{1992MNRAS.258P..41R,1538-4357-482-1-L29} and by \citet{1993ApJ...405..273W}, on the contrary, have lead to the \textit{traditional} GRB model. There, the afterglow is explained by assuming the synchrotron/synchrotron self-Compton (SSC) emission from accelerated electrons in the slowing down process of  an ultra-relativistic blastwave of $\Gamma \sim 1000$ by the circumburst medium \citep{1994ApJ...433L..85W,1997MNRAS.288L..51W,1995ApJ...455L.143S,1997ApJ...489L..37S,1998ApJ...497L..17S}. This has become known as the \textit{ultra-relativistic shockwave model}. As pointed out by \citet{2018pgrb.book.....Z}, this ultra-relativistic blastwave model has been traditionally adopted in order to explain a vast number of observations:  

\begin{enumerate}[label=(\roman*)]
    \item 
    the X-ray afterglow including the steep and the shallow decay phases all the way to the X-ray flares (see section 2.2.2 in \citealp{2018pgrb.book.....Z});

    \item the optical and the radio afterglow (see sections 2.2.3 and 2.2.4 in \citealp{2018pgrb.book.....Z});

    \item the high-energy afterglow in the GeV emission (see sections 2.2.5 in \citealp{2018pgrb.book.....Z}).
\end{enumerate}

Related to the above traditional approach were the papers by \citet{Ruffini:1975ne} and \citet{1977MNRAS.179..433B}, which addressed the gravitational accretion of magnetized plasma of infinite conductivity into a Kerr BH. Such a gravitation-dominated accretion theory implies the need of a large magnetic field ($\sim10^{15}$~G) and high density ($\sim 10^{12}$--$10^{13}$~g~cm$^{-3}$) near the last stable orbit around a $\sim 3~M_\odot$ BH. This gravitation-dominated accretion has been commonly adopted as an input for the above-mentioned ultra-relativistic jetted emission from an accretion (at a rate $\sim1~M_\odot$~s$^{-1}$) onto Kerr BH to power a GRB of luminosity $\sim10^{52}$~erg~s$^{-1}$. 

Since 2018, it has become clear that  the three above processes do not share a common origin, and they are not related to an ultra-relativistic blastwave.

An electrodynamical accretion process of ionized plasma alternative to the gravitational-dominated accretion theory, has been announced \citep[see companion paper][]{2019ApJ...883..191R}, operating at density of $\sim 10^{-14}$~g~cm$^{-3}$ (see section~\ref{sec:7}).

\subsection{Role of magnetars and spinning neutron stars}

In parallel, a variety of models have been developed adopting, instead of a BH, an energy injection from various combinations of NSs and ``magnetars''. \citet{1998A&A...333L..87D, 1998PhRvL..81.4301D,2001ApJ...552L..35Z} adopted an energy injection from a long-lasting spinning-down millisecond pulsar or a magnetar (magnetic dipole strength $\sim10^{15}$ G). Within this approach, the shallow decay or the plateau observed at times $\sim 10^2$--$10^4$~s it is attributed to the energy injection by the magnetic dipole radiation \citep[see e.g.][]{2006MNRAS.372L..19F,2007MNRAS.377.1638D,2013ApJ...779L..25F}. The magnetar model is consistent with the so-called ``internal plateaus'', namely the ones which end with a very steep decay slope, which cannot be explained solely by the external shock waves. The steep drop is thus explained by the sudden decrease of the energy injection by the pulsar/magnetar engine at the characteristic life-time of magneto-dipole emission \citep{2007ApJ...665..599T,2014ApJ...785...74L,2010MNRAS.409..531R,2013MNRAS.430.1061R,2015ApJ...805...89L,2018ApJS..236...26L}. All these alternative models converge finally to the \textit{ultra-relativistic shockwave model}. We show below how from 2018 the observations sharply constrain this model. 

As we will show below, in the binary driven hypernova (BdHN) scenario, the GRB afterglow originates from mildly-relativistic expanding SN ejecta with energy injection from the newly-born NS (hereafter $\nu$NS) at its center, and from the $\nu$NS pulsar emission itself.

\subsection{The role of binary progenitors in GRBs}

Alternatively to the above models, addressing the GRB within a single progenitor scenario, fundamental papers presented a vast number of possible binary progenitors for GRBs \citep{1999ApJ...526..152F,2003ApJ...591..288H}. Following this seminal paper, we have developed the concept of BdHN, which is recalled in Sec.~\ref{sec:2}. This model includes three different components: 1) a CO$_\mathrm{core}$ undergoing a SN explosion in presence of a binary NS companion; 2) an additional NS, indicated as a $\nu$NS, the newborn NS originating at the center of the SN, accreting the SN ejecta and giving  origin to the afterglow; 3) the formation of the BH by the hypercritical accretion of the SN ejecta onto the preexisting NS companion, giving rise to the GeV emission.

Since the beginning of 2018, there have been considerable advances in the time-resolved spectral analysis of GRBs by the state-of-the-art algorithms and tools \citep{doi:10.1063/1.1835238,2017ifs..confE.130V}. Thanks to this methodology, conceptually different from the Band function approach \citep[see e.g.][]{2019arXiv190404162R}, together with an improved feedback from three-dimensional smoothed-particle-hydrodynamics (SPH) simulations \citep{2019ApJ...871...14B}, three new results have followed from the BdHN analysis which question the traditional approach.

1) The explanation of the X-ray flares in the ``flare-plateau-afterglow'' (FPA) phase \citep{2018ApJ...852...53R} as originating from a BdHN observed in the orbital plane of the binary progenitor system. In particular, the observational data of soft X-ray flares in the early ($t\sim 100$~s rest-frame) FPA phase indicate that the emission arises from a mildly-relativistic system with Lorentz factor $\Gamma \sim 2$--$5$ \citep{2018ApJ...852...53R}. 

2) We investigated the FPA phase of GRB 130427A using the time-resolved spectral analysis on the early X-ray data \citep{2015ApJ...798...10R,2019ApJ...886...82R,2019ApJ...874...39W}. There, from the thermal emission in the FPA phase \citep[see Fig.~7 in][]{2015ApJ...798...10R}, it was established an upper limit of $\sim 0.9$~c to the expansion velocity. Such a mildly-relativistic expansion of the FPA phase emitter was further confirmed in GRB 151027A \citep{2018ApJ...869..151R} by the soft and hard X-ray observations, and in GRB 171205A by the optical emission lines \citep{2019Natur.565..324I}. It motivated the first detailed model, applied to GRB 130427A, of the plateau-afterglow emission of the FPA phase \citep{2018ApJ...869..101R,2019ApJ...874...39W}, as arising from the synchrotron radiation by relativistic electrons within the mildly-relativistic expanding SN ejecta magnetized by the $\nu$NS. 


3) One of the newest results on GRB 1901114C infers the GeV emission, originating in the traditional model at distances $10^{12}$--$10^{16}$~cm, to originate instead from electrodynamical process of BH rotational energy extraction very close to the BH horizon \citep{2019ApJ...886...82R}. This electrodynamical process occurs in a very low-density environment of $\sim 10^{-14}$~g~cm$^{-3}$, and leads to an energy per particle up to $10^{18}$~eV. This is confirmed by the simulations in the accompanying cavity generated by the BH accretion~\citep{2019ApJ...883..191R}.

All the above shows the different role in a BdHN I of three main components: the SN, the $\nu$NS and the newborn BH. In this article, we aim to further clarify, confirm and extend the explanation of the plateau-afterglow emission of the FPA phase, as powered by the SN and the $\nu$NS interaction within the BdHN scenario, following the treatment presented in \citet{2018ApJ...869..101R,2019ApJ...874...39W}. We analyze the cases of GRB 130427A, GRB 180728A, GRB 160509A, GRB 160625B and GRB 190114C.

The article is organized as follows. In Sec.~\ref{sec:2}, we recall the physical and astrophysical properties of the BdHN model. In Sec.~\ref{sec:3}, we recall the observational properties of the GRBs analyzed in this work. In Sec.~\ref{sec:4}, we simulate the X-ray afterglow of the above-mentioned sources by the mild-relativistic synchrotron model and infer the magnetic field of the $\nu$NS based on the framework presented in \citet{2019ApJ...874...39W}. The nature of the obtained magnetic field of the $\nu$NS is discussed in Sec.~\ref{sec:5}. In Sec.~\ref{sec:6}, we discuss the possible nature of the magnetic field around the newborn BH in a BdHN. Finally, in Sec.~\ref{sec:7} we outline our conclusions.

\section{The binary-driven hypernova (BdHN) scenario}\label{sec:2}

\begin{figure*}
    \centering
    \includegraphics[width=0.8\hsize,height=17cm]{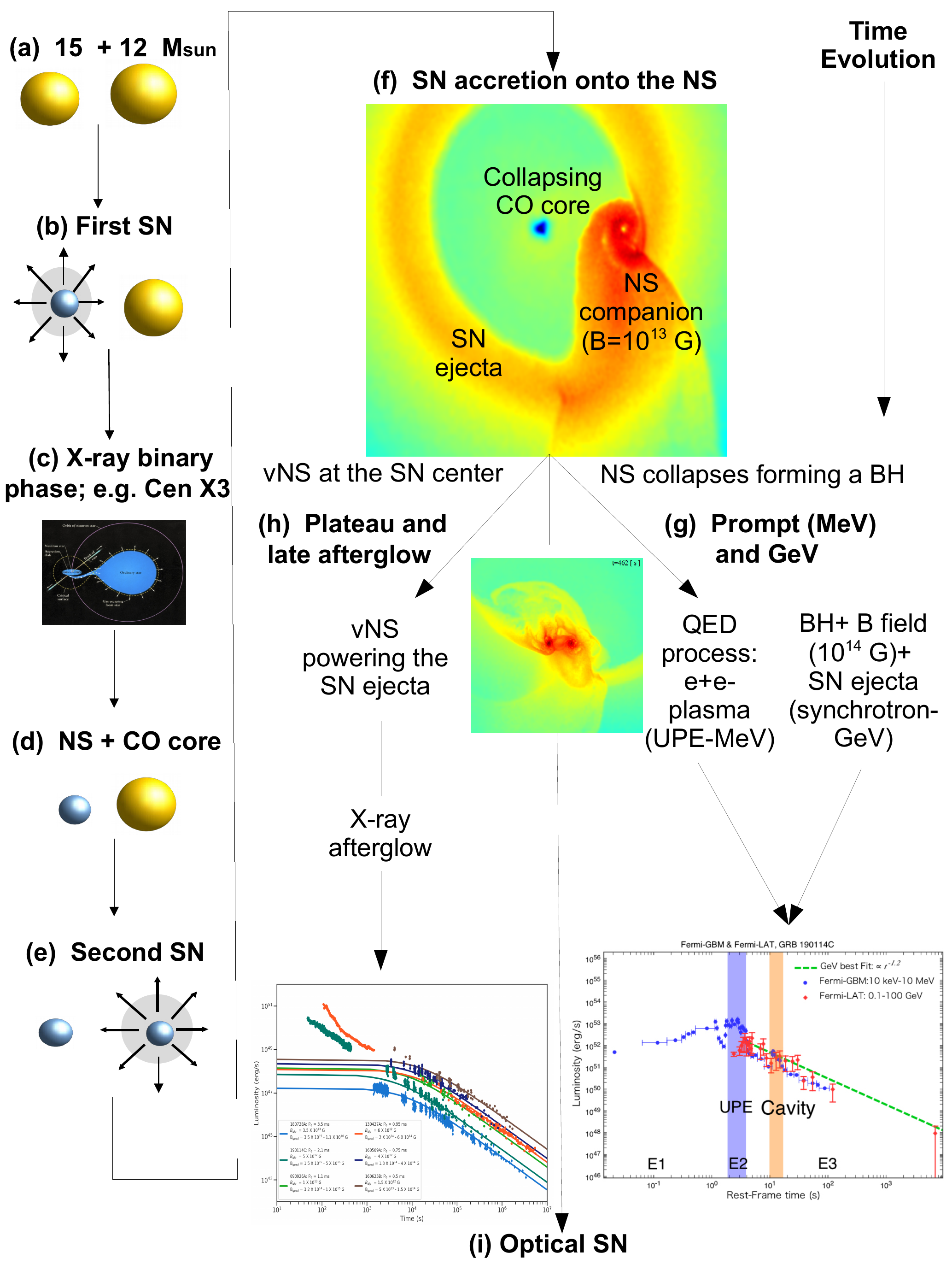}
    \caption{Schematic evolutionary path of a massive binary up to the emission of a BdHN. (a) Binary system composed of two main-sequence stars of $15$ and $12$ solar masses, respectively. (b) At a given time, the more massive star undergoes the core-collapse SN and forms a NS (which might have a magnetic field $B\sim 10^{13}$~G). (c) The system enters the X-ray binary phase. (d) The core of the remaining evolved star, rich in carbon and oxygen, for short CO$_{\rm core}$, is left exposed since the hydrogen and helium envelope have been striped by binary interactions and possibly multiple common-envelope phases (not shown in this diagram). The system is, at this stage, a CO$_{\rm core}$-NS binary, which is taken as the initial configuration of the BdHN model \citep{2014ApJ...793L..36F,2015ApJ...812..100B,2016ApJ...833..107B,2019ApJ...871...14B}. (e) The CO$_{\rm core}$ explodes as SN when the binary period is of the order of few minutes, the SN ejecta of a few solar masses start to expand and a fast rotating, newborn NS, for short $\nu$NS, is left in the center. (f) The SN ejecta accrete onto the NS companion, forming a massive NS (BdHN II) or a BH (BdHN I; this example), depending on the initial NS mass and the binary separation. Conservation of magnetic flux and possibly additional MHD processes amplify the magnetic field from the NS value to $B~\sim 10^{14}$~G around the newborn BH. At this stage the system is a $\nu$NS-BH binary surrounded by ionized matter of the expanding ejecta. (g) The accretion, the formation and the activities of the BH contribute to the GRB prompt gamma-ray emission and GeV emission (not the topic of this work)}.
    \label{fig:BdHN}
\end{figure*}

The BdHN model has been introduced for the explanation of long-duration gamma-ray bursts (GRBs) and it is based on the induced gravitational collapse (IGC) paradigm \citep{2012ApJ...758L...7R}, occurring in a specific binary system which follows from a specific evolutionary path \citep[see  Fig.~\ref{fig:BdHN} and][for details]{2014ApJ...793L..36F,2015ApJ...812..100B,2015PhRvL.115w1102F,2019Univ....5..110R}.

{
As Fig.~\ref{fig:BdHN} shows, the system starts with a binary composed of two main-sequence stars, say of $15$ and $12$ solar masses, respectively. At a given time, at the end of its thermonuclear evolution, the more massive star undergoes the core-collapse supernova (SN) and forms a neutron star (NS). The system then enters the X-ray binary phase. After possibly multiple common-envelope phases and binary interactions \citep[see][and references therein]{2014ApJ...793L..36F,2015PhRvL.115w1102F}, the hydrogen and helium envelope of the other main-sequence star are stripped, leaving exposed its core that is rich in carbon and oxygen. For short, we refer to it as carbon-oxygen core (CO$_{\rm core}$) following the literature on the subject \citep[see e.g.][]{1994Natur.371..227N,1995ApJ...450L..11F,2000ApJ...534..660I,2006Natur.442.1011P,2011MNRAS.412L..78Y}. The system at this stage is a CO$_{\rm core}$-NS binary in tight orbit (period of the order of few minutes), which is taken as the initial configuration of the BdHN scenario in which the IGC phenomenon occurs \citep{2014ApJ...793L..36F,2015ApJ...812..100B,2016ApJ...833..107B,2019ApJ...871...14B}.
}

{We now proceed to describe the BdHN scenario. At the end of its thermonuclear evolution the CO$_{\rm core}$ undergoes a core-collapse SN (of type Ic in view of the hydrogen and helium absence). Matter is ejected but also a the center of the SN, a newborn NS is formed, for short referred to as $\nu$NS, to differentiate it from the accreting NS binary companion. As we shall see, this differentiation is necessary in view of the physical phenomena and corresponding observables in a BdHN associated with each of them.} Owing to the short orbital period, the SN ejecta produce a hypercritical (i.e. highly super-Eddington) accretion process onto the NS companion. The material hits the NS surface developing and outward shock which creates an accretion ``atmosphere'' of very high density and temperature on top the NS. These conditions turn to be appropriate for the thermal production of positron-electron ($e^+e^-$) pairs which, when annihilating, leads to a copious production of neutrino-antineutrino ($\nu\bar\nu$) which turn to be the most important carriers of the gravitational energy gain of the accreting matter, allowing the rapid and massive accretion to continue. We refer to \citet{2014ApJ...793L..36F,2016ApJ...833..107B,2018ApJ...852..120B} for details on the hypercritical accretion and the involved neutrino physics.

Depending on the specific system parameters, i.e. mass of the binary components, orbital period, SN explosion energy, etc, two possible fates for the NS are possible {(see \citet{2015ApJ...812..100B,2016ApJ...833..107B,2019ApJ...871...14B} for details on the relative influence of each parameter in the system)}. For short binary periods, i.e.~$\sim 5$~min, the NS reaches the critical mass for gravitational collapse and forms a BH \citep[see e.g.][]{2015PhRvL.115w1102F,2015ApJ...812..100B,2016ApJ...833..107B,2019ApJ...871...14B}. We have called this kind of system a BdHN type I \citep{2019ApJ...874...39W}. A BdHN I emits an isotropic energy $E_{\rm iso}\gtrsim 10^{52}$~erg and gives origin to a new binary composed by the NS formed at the center of the SN, hereafter $\nu$NS, and the BH formed by the collapse of the NS. For longer binary periods, the hypercritical accretion onto the NS is not sufficient to bring it to the critical mass and a more massive NS (MNS) is formed. We have called these systems BdHN of type II \citep{2019ApJ...874...39W} and they emit energies $E_{\rm iso}\lesssim 10^{52}$~erg. A BdHN II gives origin to a new binary composed by the $\nu$NS and the MNS.

The BdHNe I represent, in our binary classification of GRBs, the totality of long GRBs with energy larger than $10^{52}$~erg while, the BdHN II with their energy smaller than $10^{52}$~erg, are far from unique and there is a variety of long GRBs in addition to them which can have similar energetics; e.g. double white dwarf (WD-WD) mergers and NS-WD mergers \citep[see][for details]{2016ApJ...832..136R,2018ApJ...859...30R,2019ApJ...874...39W}.
\begin{figure*}
\centering
\includegraphics[scale=0.33]{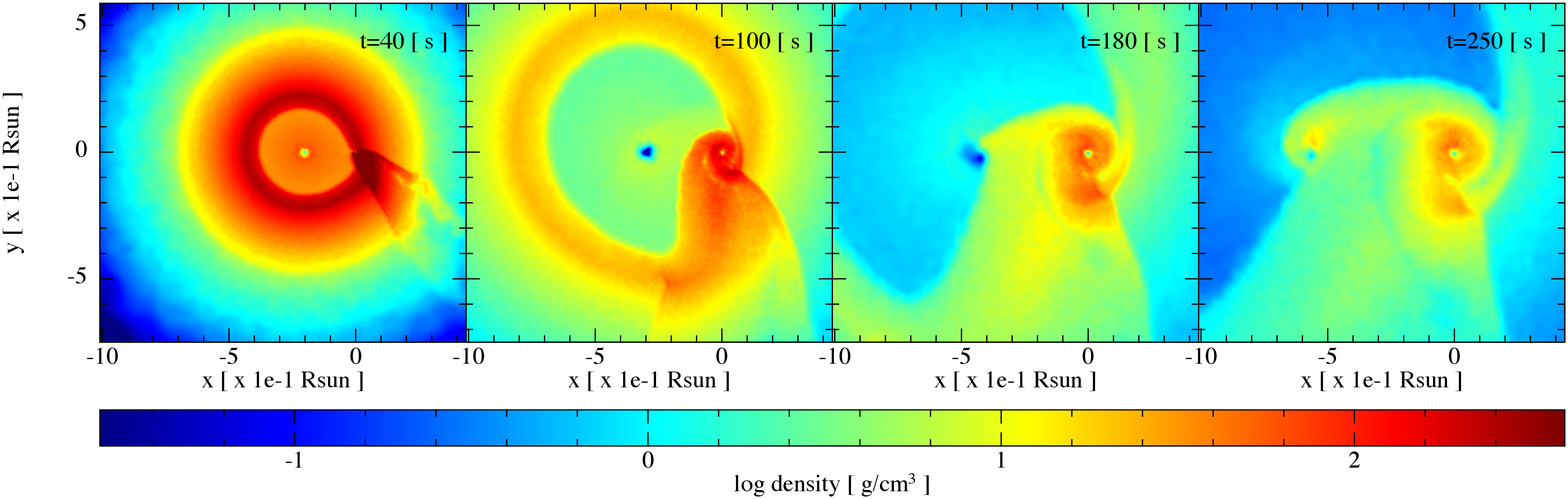}
\includegraphics[scale=0.33]{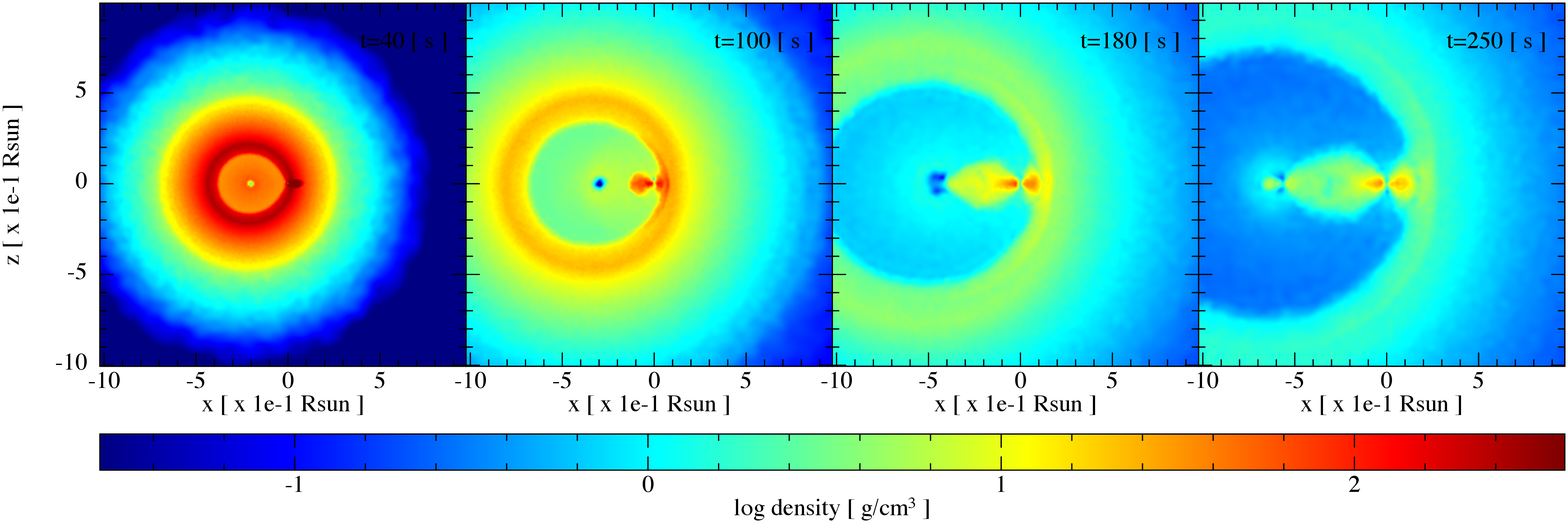}
\caption{SPH simulation of a BdHN I: model `25M1p1e' of Table~$2$ in \citet{2019ApJ...871...14B}. The binary progenitor is composed of a CO$_{\rm core}$ of $\approx 7~M_\odot$ produced by a zero-age main-sequence star (ZAMS) star of $25\,M_\odot$ (see Table~1 in \citealp{2019ApJ...871...14B}), and a $2\,M_\odot$ NS companion. The orbital period is $\approx 5$~min. Each frame, from left to right, corresponds to selected increasing times being $t=0$~s the instant of the SN shock breakout. The upper panel shows the mass density on the equatorial plane and the lower panel the plane orthogonal to the equatorial one. The reference system is rotated and translated to align the x-axis with the line joining the binary components. The origin of the reference system is located at the NS companion position. The first frame corresponds to $t=40$~s and it shows that the particles entered into the NS capture region forms a tail behind it. These particles then circularize around the NS forming a thick disk which is already visible in the second frame at $t=100$~s. Part of the SN ejecta is also attracted by the $\nu$NS accreting onto it; this is appreciable in the third frame at $t=180$~s. At $t=250$~s (about one orbital period), a disk structure has been formed around the $\nu$NS and the NS companion. To guide the eye, the $\nu$NS is at the x-coordinate: $-2.02$, $-2.92$, $-3.73$ and $-5.64$ for $t=40$~s, $100$~s, $180$~s and $250$~s, respectively. This figure has been produced with the \emph{SNsplash} visualization program \citep{2011ascl.soft03004P}. The figure has been taken from \citet{2019ApJ...871...14B} with the permission of the authors.}
\label{fig:25Mzams2Mns}
\end{figure*}

Three-dimensional, numerical SPH simulations of BdHNe have been recently presented in \citet{2019ApJ...871...14B}. These simulations improve and extends the previous ones by \citet{2016ApJ...833..107B}. A fundamental contribution of these simulations has been to provide a visualisation of the morphology of the SN ejecta which is modified from the initial spherical symmetry. A low-density cavity is carved by the NS companion and, once its collapses, further by the BH formation process \citep[see also][]{2019arXiv190404162R}. Such an asymmetric density distribution leads to a dependence of the GRB description as a function of the observer viewing angle: in the orbital/equatorial plane or in the plane orthogonal to it \citep{2016ApJ...833..107B,2018ApJ...852...53R,2018ApJ...869..151R,2019ApJ...871...14B} and as a function of the orbital period of the binary, in the simulation of Fig.~\ref{fig:25Mzams2Mns} about $300$~s \citep{2018ApJ...869..151R}.

The SN transforms into a hypernova (HN) as a result of the energy and momentum transfer of the $e^+e^-$ plasma \citep{2018ApJ...869..151R,2019ApJ...871...14B}. The SN shock breakout and the hypercritical accretion can be observed as X-ray precursors \citep{2016ApJ...833..107B,2019ApJ...874...39W}. The $e^+e^-$ feedback also produces gamma- and X-ray flares observed in the early afterglow \citep{2018ApJ...852...53R}. There is then the most interesting emission episode which is related to the $\nu$NS originated from the SN explosion. Namely, the synchrotron emission by relativistic electrons, injected from the $\nu$NS pulsar emission into the HN ejecta in presence of the $\nu$NS magnetic field, explain the X-ray afterglow and its power-law luminosity \citep{2018ApJ...869..101R,2019ApJ...874...39W}. Finally, the HN is observed in the optical bands few days after the GRB trigger, powered by the energy release of the nickel decay. 

Figure~\ref{fig:BdHN} and Table~\ref{tab:observables} summarize the above correspondence between the BdHN physical process and each GRB observable, emphasizing the role of each component of the binary system. We also refer the reader to \citet{2019Univ....5..110R}, and references therein, for a recent review on the physical processes at work and related observables in BdHNe I and II.

\begin{table*}
    \centering
    \caption{{Summary of the GRB observables associated with each BdHN I component and physical phenomena.}}
    {
    \begin{tabular}{l|c|c|c|c|c}
    \hline
     BdHN component/phenomena &  \multicolumn{5}{c}{GRB observable} \\
     \cline{1-6}
          & X-ray & Prompt & GeV-TeV & X-ray flares & X-ray plateau\\
          &    precursor      & (MeV) & emission & early afterglow & and late afterglow\\
          \cline{1-6}
    SN breakout$^a$ & $\bigotimes$ & & & & \\
    \cline{1-1}
    Hypercritical accretion onto the NS$^b$  &  $\bigotimes$ & & & & \\
      \cline{1-1}
    $e^+e^-$ from BH formation: transparency & & $\bigotimes$ & & &\\
    in low baryon load region$^c$  & & & & & \\
     \cline{1-1}
    \textit{Inner engine}: newborn BH + $B$-field+SN ejecta$^d$  & & & $\bigotimes$ & &\\
     \cline{1-1}
    $e^+e^-$ from BH formation: transparency & & & & $\bigotimes$ &\\
    in high baryon load region (SN ejecta)$^e$  & & & & &\\
     \cline{1-1}
    Synchrotron emission by $\nu$NS injected & & & & & $\bigotimes$\\
    particles on SN ejecta$^f$ & & & & & \\
     \cline{1-1}
    $\nu$NS pulsar-like emission$^f$ & & & & & $\bigotimes$\\
    \cline{1-6}
    \end{tabular}
    }
    \tablerefs{$^a$\citet{2019ApJ...874...39W},$^b$\citet{2014ApJ...793L..36F,2016ApJ...833..107B,2019Univ....5..110R},$^c$\citet{2001A&A...368..377B},$^d$\citet{2018arXiv181101839R,2019ApJ...886...82R,2019arXiv190404162R,2019ApJ...883..191R}, $^e$\citet{2018ApJ...852...53R}, $^f$\citet{2018ApJ...869..101R,2019ApJ...874...39W} and this work.}
    \label{tab:observables}
\end{table*}

\section{GRBs (BdHNe I) of the present work}\label{sec:3}

\noindent\textbf{GRB 130427A} is one of the best observed GRBs, it locates at redshift $z\sim 0.34$ \citep{2013GCN..14455...1L}, more than 50 observatories participated the observation. It hits the record of the brightness in the gamma-ray emission, so that Fermi-GBM was saturated. It also hits the record of GeV observation with more that 500 photons above $100$~MeV received, and the GeV emission observed till $\sim 10^4$~s \citep{2014Sci...343...42A}. 

The shape of its prompt emission consists a $\sim 3$~s precursor, followed by a multipeaked pulse lasting $\sim 10$~s. At time $\sim 120$~s, an additional flare appears, then it enters the afterglow \citep{2014Sci...343...48M}. The X-ray afterglow is observed by Swift and NuStar. Swift covers discretely from $\sim 150$~s to $\sim 10^7$~s \citep{2015ApJ...805...13L}, and NuStar observes three epochs, starting approximately $1.2$, $4.8$ and $5.4$ days, for observational duration $30.5$, $21.2$, and $12.3$ ks \citep{2013ApJ...779L...1K}. The power-law decay index of the late time afterglow after $\sim 2000$~s gives $\sim -1.32$ \citep{2015ApJ...798...10R}.

The optical spectrum reveals that $16.7$ days after the GRB trigger, a typical of SNe Ic emerges \citep{2013ApJ...776...98X, 2018ApJS..234...26L}, as predicted by \citet{2013GCN..14526...1R}.

\vspace{1em}
\noindent\textbf{GRB 160509A}, at redshift $z\sim 1.17$ \citep{2016GCN.19419....1T}, is a strong source of GeV emission, including a $52$ GeV photon arriving at $77$~s, and a $29$ GeV photon arriving $\sim 70$~ks \citep{2016ApJ...833...88L}.

GRB 160509A consists of two emission periods, $0-40$~s and $280-420$s \citep{2017ApJ...844L...7T}. The first period exhibits a single pulse structure for sub-MeV emission, and a double pulses structure for $\sim 100$~MeV emission. The second period is in the sub-MeV energy range with double pulses structure. Swift-XRT started the observation $\sim 7000$~s after the burst, with a shallow power-law decay of index $\sim -0.6$, followed by a normal decay of power-law index $\sim -1.45$ after $5\times 10^4$~s \citep{2017ApJ...844L...7T,2018ApJS..236...26L}.

There is no supernova association reported, the optical signal of supernova can hardly be confirmed for GRBs with redshift $>1$, since the absorption is intense \citep{2006ARA&A..44..507W}.

\vspace{1em}
\noindent\textbf{GRB 160625B}, at redshift $1.406$ \citep{2016GCN.19600....1X}, is a bright GRB with the speciality that the polarisation has been detected. Fermi-LAT has detected more than 300 photons with energy  $>100$~ MeV \citep{2017ApJ...849...71L}. 

The gamma-ray light curve has three distinct pulses \citep{2019ApJS..242...16L,2018NatAs...2...69Z}. The
first short pulse is totally thermal, it lasts $\sim 2$~s; the second bright pulse starts from $\sim 180$~s and ends at $\sim 240$~s; the last weak pulse emerges from $\sim 330$~s and lasts $\sim 300$~s. The total isotropic energy reaches $\sim 3 \times 10^{54}$~erg \citep{2017ApJ...848...69A,2017ApJ...849...71L}.

Swift-XRT starts the observation at late time ($> 10^4$~s), a power-law behaviour with decaying index $\sim -1.25$. 

There is no supernova confirmation, possibly it is due to the redshift $>1$ \citep{2006ARA&A..44..507W}.

\vspace{1em}
\noindent\textbf{GRB 190114C}, at redshift $z\sim 0.42$ \citep{2019GCN.23695....1S}, is the first GRB with TeV photon detection by MAGIC \citep{2019GCN.23701....1M,2019Natur.575..455M}. It has twin features as GRB 130427A \citep{2019arXiv190107505W}, and it caught great attention as well.

The prompt emission of GRB 190114C starts by a multipeaked pulse, its initial $\sim 1.5$~s is non-thermal, then followed by a possible thermal emission till $\sim 1.8$~s. The confident thermal emission exists during the peak of the pulse, from $2.7 - 5.5$~s. The GeV emission starts from $2.7$~s, initiated with a spiky structure, then follows a power-law decay with index $\sim -1.2$ \citep{2019arXiv190404162R}. The GeV emission is very luminous, more than $200$ photons with energy $>100$~MeV are received. The X-ray afterglow observed by Swift-XRT shows a persistent power-law decay behaviour, with decaying index $\sim 1.35$ \citep{2019arXiv190107505W}.

An continuous observational campaign lasting $\sim 50$ days unveiled the SN emergence at $\sim 15$~days after the GRB \citep{2019GCN.23983....1G}, which is consistent with the prediction of $18.8 \pm 3.7$ days after the GRB by \citet{2019GCN.23715....1R}.

\section{X-ray afterglow of GRB and Magnetic field of $\nu$NS}\label{sec:4}

The newborn NS at the center of the SN, i.e the $\nu$NS, ejects high-energy particles as in traditional pulsar models. This means that these particles escape from the $\nu$NS magnetosphere through the so-called ``open'' magnetic field lines, namely the field lines which cannot close within the light cylinder radius that determines the size of the co-rotating magnetosphere. Those particles interact with the SN ejecta, which by expanding in the $\nu$NS magnetic field, produce synchrotron radiation which we discuss below. Hence, the acceleration mechanism is similar to the one occurring in traditional SN remnants but with two main differences in our case: 1) we have a $\sim 1$~ms $\nu$NS pulsar powering the SN ejecta and 2) the SN ejecta are at a radius $\sim 10^{12}$~cm at the beginning of the afterglow, at rest-frame time $t\sim 100$~s, since the SN expands with velocity $\sim 0.1\,c$.

The above distance is well beyond the light cylinder radius, so it is expected that only the toroidal component of the magnetic field, which decreases as $1/r$ (see Eqs.~\ref{eq:magneticField} and \ref{eq:BtBs}), survives \citep[see, e.g.,][for details]{1969ApJ...157..869G}. Therefore, the relevant magnetic field for the synchrotron radiation in the afterglow is the one of the $\nu$NS which is stronger (as shown below at that distance is of the order of $10^5$~G) that the one possibly produced inside the remnant by dilute plasma currents, unlike the traditional models for the emission of old ($\gtrsim 1$~kyr) SN remnants.

In \citet{2018ApJ...869..101R} and \citet{2019ApJ...874...39W}, we simulate the afterglow by the synchrotron emission of electrons from the optically thin region of the SN ejecta, that expands mildly-relativistic in the $\nu$NS magnetic field. The FPA emission at times $t\gtrsim 10^2$~s has two origins: the emission before the plateau phase ($\sim 5\times10^3$~s) is mainly contributed by the remaining kinetic energy of the SN ejecta, and at later times, the continuous energy injection from the $\nu$NS takes over the dominance. We extend the same approach in this paper to the GRBs of section \ref{sec:3}.




To fully follow the temporal behaviour of radiation spectra, it is necessary to solve the kinetic equation for the electron distribution in the transparent region of the SN ejecta:
\begin{equation}\label{eq:kinetic_eq}
    \frac{\partial N(\gamma,t)}{\partial t} = \frac{\partial}{\partial \gamma} (\dot{\gamma} (\gamma,t)N(\gamma,t))+Q(\gamma,t) \, ,
\end{equation}
where $N(\gamma,t)$ is the electron number distribution as a function of electron energy $\gamma=E/m_e c^2$, $\dot{\gamma} (\gamma,t)$ is the electron energy loss rate normalized to the electron rest-mass, $ Q(\gamma,t) = Q_0(t)\gamma^{-p}$ is the particle injection rate, assumed to be a the power-law of index $p$, so the electrons injected are within the energy range of $\gamma_{\min}$ and $\gamma_{\max}$. The total injection luminosity $L_{\text{inj}}(t)$ is provided by the kinetic energy of the SN and the rotational energy of the $\nu$NS, here parametrized via the power-law injection power
\begin{equation}\label{eq:injection_power}
     L_{\text{inj}}(t) = \int\limits_{\gamma_{\min}}^{\gamma_{\max}} Q(\gamma,t)d\gamma \simeq L_0 \left(1+\frac{t}{\tau_0}\right)^{-k},
\end{equation}
where $L_0$, $k$ and $\tau_0$ are assessed by fitting of the light curve data. The major energy loss is considered as the adiabatic energy loss and the synchrotron energy loss

%
%

%
%
%

\begin{equation}
    \label{eq:losses}
   \dot{\gamma} (\gamma,t) = \frac{\dot{R}(t)}{R(t)}\gamma + \frac{4}{3}\frac{\sigma_T}{m_e c}\frac{B(t)^2}{8\pi}\gamma^2 \, ,
\end{equation}
where $R(t)$ is the size of emitter, $\sigma_T$ is the Thomson cross section and $B(t)$ is the magnetic field strength expected to have toroidal configuration given by
\begin{equation}
    B(t) = B_0 \left(\frac{R(t)}{R_0}\right)^{-1} \, ,
    \label{eq:magneticField}
\end{equation}
here $B_0$ is the magnetic field strength at the distance $R_0$. The final bolometric synchrotron luminosity from this system gives
\begin{equation}
    \label{eq:synspec}
    L_\mathrm{syn}(\nu,t) = \int\limits_1^{\gamma_{\max}} N(\gamma,t) P_\mathrm{syn}(\nu,\gamma,B(t)) d\gamma\, .
\end{equation}


As we have introduced in section \ref{sec:1}, the thermal emission during the FPA phase indicates a mildly-relativistic velocity $\sim  0.5-0.9~c$ at time $\sim 100$~s \citep{2015ApJ...798...10R,2018ApJ...852...53R,2019ApJ...886...82R,2019ApJ...874...39W}. We adopt this value as the initial velocity and radius of the transparent part of SN ejecta.

For later stages at around $10^6$~s, when a sizable front shell of SN ejecta becomes transparent, we adopt the velocity of $\sim 0.1~c$ obtained through observations of Fe II emission lines~\citep[see e.g.][]{2013ApJ...776...98X}. We make the simplest assumption of a uniformly decelerating expansion during the time interval $10^2\lesssim t\lesssim 10^6$~s. The SN ejecta remains in the coasting phase for hundreds of years \citep[see e.g.][]{1997ApJ...490..619S}, therefore, we adopt a constant velocity from $10^6$~s till $10^7$~s.

Following the above discussion and our data analysis, we describe the expansion velocity, as
\begin{equation}
    \label{eq:expansion}
	\dot{R}(t) =  
	\begin{cases} 
	v_0 - a_0 \, t & 10^2 <  t < 10^6 \mathrm{s} \\
	v_f & 10^7 > t > 10^6 \mathrm{s} 
	\end{cases} \, 
\end{equation}
with typical value $v_0 = 2.4 \times 10^{10}$~cm~s$^{-1}$, $a_0=2.1 \times 10^{4}$~cm~s$^{-2}$ and $v_f = 3 \times 10^{9}$~cm~s$^{-1}$.

It is appropriate to clarify how the model parameters presented in this table are obtained: $R_0$ and $\tau_0$ are fixed by the observed thermal component at around $10^2$~s, from which we obtain the radius and expansion velocity of the SN front. The minimum and maximum energy of the injected electrons, $\gamma_\text{min}$ and $\gamma_\text{max}$, are fixed once $B$ is given. $L_0$ is fixed by a normalization of the observed source luminosity. The power-law index of the energy injection rate, $p$, is fixed to the value $p=3/2$. The parameter $k$ is fixed to produce the power-law decay of the late time X-ray data. Therefore, the ``free parameter'' to be obtained is $B_0$.

\begin{figure}
	\centering
	\includegraphics[width=\hsize,clip]{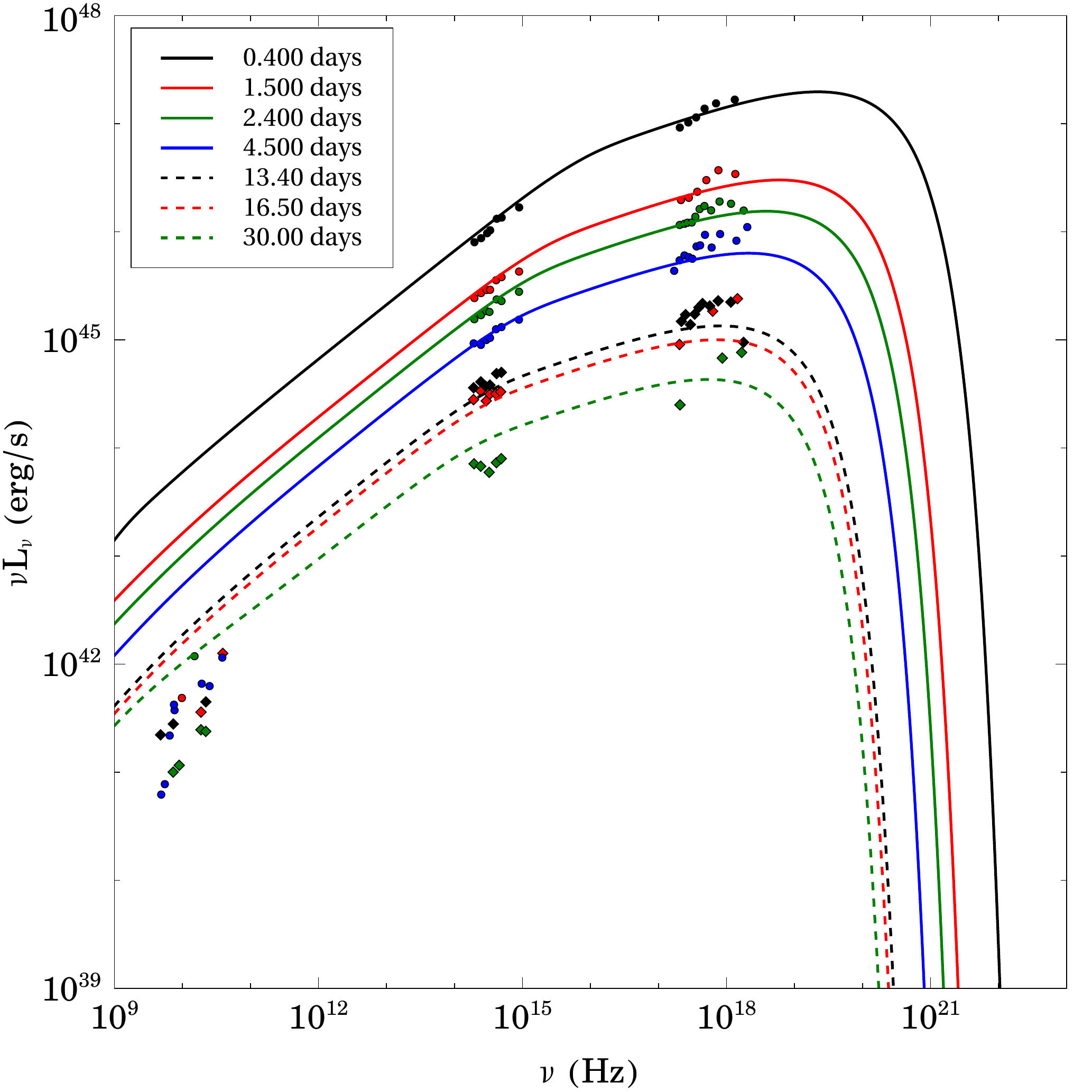}
	\caption{Model evolution of synchrotron spectral luminosity at various times compared with measurements in various spectral bands for GRB 160625B.}
	\label{fig:GRB_fit_160625B}
\end{figure}

In \citet{2018ApJ...869..101R}, we have given detailed fitting parameters and figures of GRB 130427A. In this article, we additionally fit GRB 160625B and confirm that the mildly relativistic model is capable of producing GRBs afteglow. As it is shown in Table~\ref{tab:parameters2} and Fig.~\ref{fig:GRB_fit_160625B}, our model fits very well the optical and the X-ray spectrum but not the GeV data. This is in agreement with the BdHN paradigm since the GeV emission is expected to be explained from the newborn BH activity and not from the $\nu$NS one \citep{2019ApJ...886...82R}. On the other hand, radio data show lack of expected flux which comes from synchrotron self-absorption processes which are rather complicated to model in current numerical framework but can be thoroughly neglected at frequencies above $10^{14}$~Hz.

\begin{table}
    \centering
    \caption{Parameters used for simulation of GRB 160625B.\label{tab:parameters2}}
    \begin{tabular}{c|c} 
    \hline
    Parameter & Value\\
    \hline
    $B_0$ & $1.0 \times 10^6 \; \mathrm{G}$ \\
    $R_0$ & $1.2\times 10^{11}\; \mathrm{cm}$ \\
    $L_{0}$ & $8.44 \times10^{52} \; \mathrm{erg/s}$  \\
    $k$ & $1.42$ \\
    $\tau_0$ & $5.0 \times 10^0 \; \mathrm{s}$\\ 
    $p$ & $1.5$\\
    $\gamma_\mathrm{min}$ & $4.0 \times 10^3$\\
    $\gamma_\mathrm{max}$ & $1.0 \times 10^6$\\
    \hline
    \end{tabular}
\end{table}
    
\begin{figure*}
	\centering
	\includegraphics[width=0.7\hsize,clip]{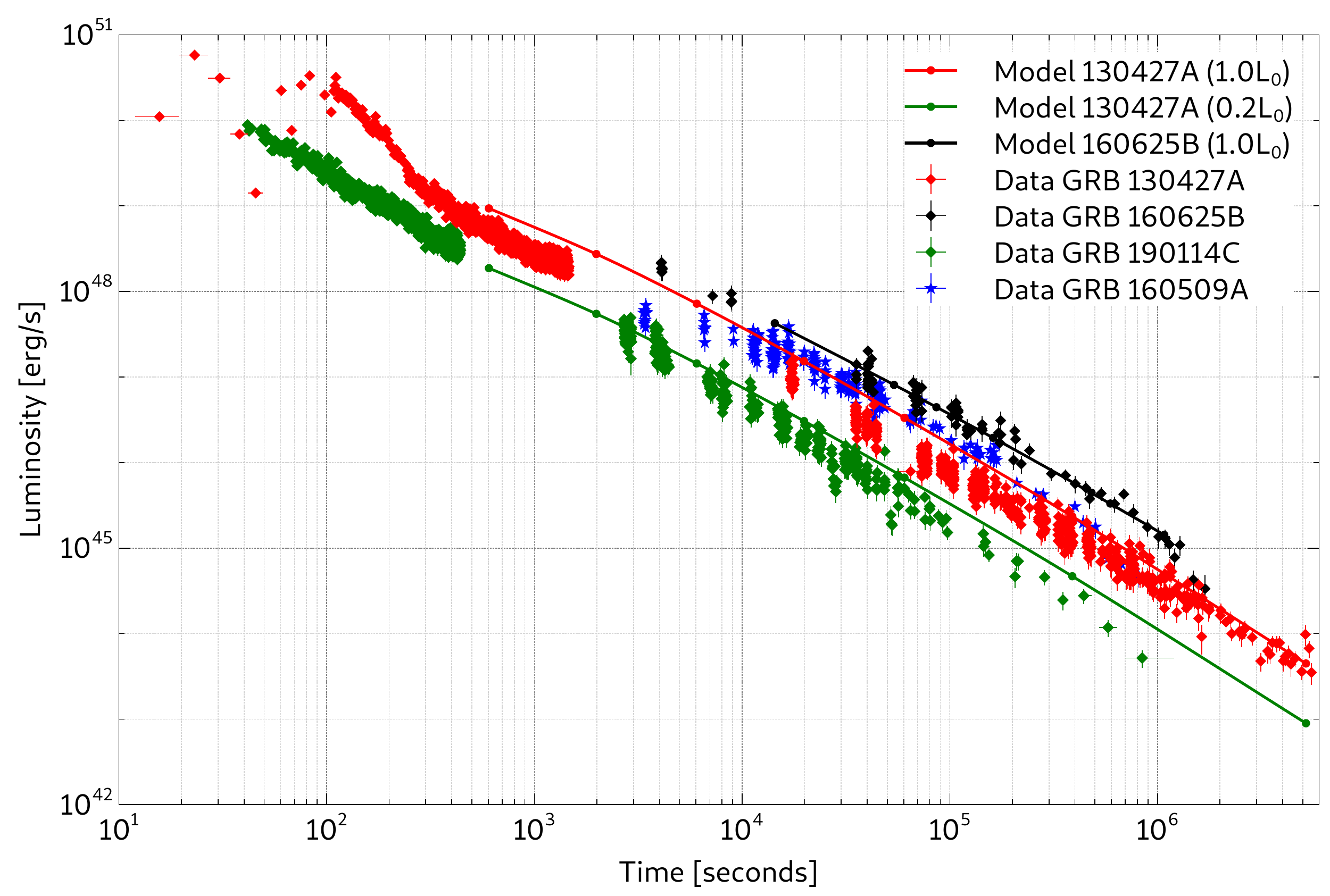}
	\caption{X-ray light-curve of GRB 160625B, GRB 130427A, GRB 190114C and GRB 160509A (black, red and green diamonds and blue stars with error bars respectively). Simulated synchrotron light curves in Swift X-ray band are shown for GRB 160625B (black line) and GRB 130427A (red line). It is also shown how, by scaling the injection power by a factor 1/5, the light curve of GRB 130427A scales down (from the red line to the green one) fitting the data of GRB 190114C}.
	\label{fig:lum_comparison}
\end{figure*}

Comparing the fitting parameters, GRB 130427A and GRB 160625B are similar except the constant of injection power $L_0$ (see Eq.~\ref{eq:injection_power}). Such similarities can be extented to other ones. It can be seen from Fig.~\ref{fig:lum_comparison}, that taking everything else as similar, from magnetic field strength and structure to expansion evolution, the GRB 190114C simulated light curve at the relevant times, can be obtained from the one of GRB 130427A, by scaling $L_0$ a factor of $1/5$.

The injection power index $k \sim 1.5$ from the fitting suggests that the quadruple emission from a pulsar dominates the late-time afterglow. As we will see below, the complementary analysis allows to infer the initial rotation period of the $\nu$NS as well as an independent estimate of its magnetic field structure.

Being just born, the $\nu$NS must be rapidly rotating and as such it contains abundant rotational energy:
\begin{equation}
	E = \frac{1}{2} I \Omega^2,
\end{equation}
where $I$ is the moment of inertia, and $\Omega = 2\pi/P_{\nu\rm NS}$ is the angular velocity. For a millisecond $\nu$NS and $I \sim 10^{45}$~g cm$^2$, the total rotational energy $E \sim 2 \times 10^{52}$~erg. Assuming that the rotational energy loss is driven by magnetic dipole and quadruple radiation we have:
\begin{multline}
	L_{\text{NS}}(t) = \frac{dE}{dt} = -I \Omega  \dot{\Omega } \\
    = - \frac{2}{3 c^3} \Omega^4 B_{\rm dip}^2 R_{\nu\rm NS}^6 \sin^2\chi_1  \left(1+\eta^2 \frac{16}{45} \frac{R_{\nu\rm NS}^2 \Omega^2}{c^2}\right),
\end{multline}\label{eq:pulsar_luminosity}
where
\begin{equation}
    \eta^2 = (\cos^2\chi_2+10\sin^2\chi_2) \frac{B_{\rm quad}^2}{B_{\rm dip}^2},
\label{eq:eta}
\end{equation}
with $\chi_1$ and $\chi_2$ the inclination angles of the magnetic moment, $B_{\rm dip}$ and $B_{\rm quad}$ are the dipole and quadruple magnetic field, respectively. The parameter $\eta$ measures the quadruple to dipole magnetic field strength ratio.

\begin{figure*}
\centering
\includegraphics[width=0.6\hsize,clip,angle=-90]{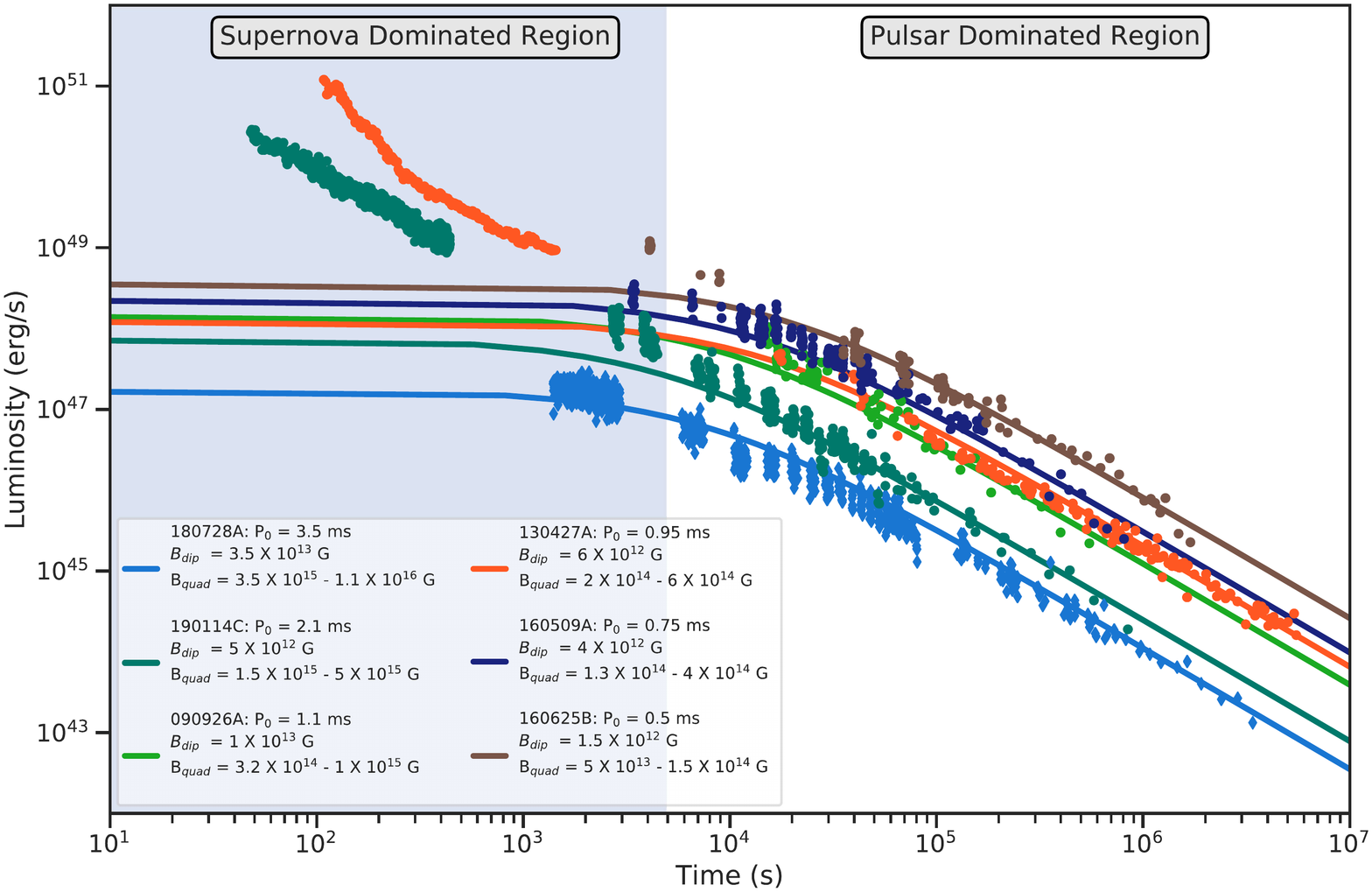}
\caption{The brown, deep blue, orange, green and bright blue points correspond to the bolometric (about $\sim 5$ times brighter than the soft X-ray observed by Swift-XRT inferred from the fitted sychrotron spectrum) light curves of GRB 160625B, 160509A, 130427A, 190114C and 180728A, respectively. The lines are the fitting of the energy injection from the rotational energy of the pulsar. The pulsar powers the late afterglow ($t \gtrsim 5 \times 10^3$~s, white background), while in the earlier time ($t \lesssim 5 \times 10^3$~s, dusty blue background), the remaining kinectic energy of the SN ejecta plays the leading role.} The fitted parameters are shown in the legend and in Table \ref{tab:ns_parameters}, the quadruple field are given in a range, its upper value is $3$ times the lower value, this is due to the oscillation angle $\chi_2$, which is a free parameter. The fittings of GRB 1340427A and 180728A are reproduced from \citet{2019ApJ...874...39W}.
\label{fig:lightcurves}
\end{figure*}

Figure~\ref{fig:lightcurves} shows the bolometric light curves ($\sim 5$ times brighter than the Swift-XRT light curves inferred from the fitting)  of GRB 160625B, 160509A, 130427A, 190114C and 180728A, respectively. We show the $\nu$NS luminosities $L_{\text{NS}}(t)$ fit the light curves. We report the fitting $\nu$NS parameters: dipole ($B_{\rm dip}$) and quadrupole ($B_{\rm quad}$) magnetic field component, initial rotation period ($P_{\nu \rm NS}$) and assuming a $\nu$NS of mass and radius, respectively, $1.4 M_\odot$ and $10^{6}$~cm. The results are also summarized in Table~\ref{tab:ns_parameters}. It becomes also clear from this analysis that the solely $\nu$NS emission is not able to explain the emission of the FPA phase at early times $10^2$--$10^3$~s. As we have shown, that emission is mainly powered by the mildly-relativistic SN kinetic energy.

\begin{table*}[!ht]
\centering
\caption{Observational properties of the GRB and inferred physical quantities of the $\nu$NS of the corresponding BdHN model that fits the GRB data. Column 1: GRB name; column 2: identified BdHN type; column 3: the isotropic energy released ($E_{\rm iso}$) in gamma-rays; column 4: cosmological redshift ($z$); column 5: $\nu$NS rotation period ($P_{\nu \rm NS}$), column 6: $\nu$NS rotational energy ($E_{\rm rot}$); columns 7 and 8: strength of the dipole ($B_{\rm dip}$) and quadrupole ($B_{\rm quad}$) magnetic field components of the $\nu$NS. The quadruple magnetic field component is given in a range that the upper limit is three times than the lower limit, this is brought by the freedom of inclination angles of the magnetic moment.  During the fitting, we consistently assume the NS mass of $1.4 M_\odot$ and the NS radius of $10^{6}$~cm for all these three cases. The fitted light-curves are shown in Fig.~\ref{fig:lightcurves}, the parameters of GRB 1340427A and 180728A are taken from \citet{2019ApJ...874...39W}.}
\label{tab:ns_parameters}
\small
\begin{tabular}{@{}cccccccc@{}}
\hline
GRB     & Type    & Redshift & $E_{\rm iso}$  & $P_{\nu \rm NS}$  & $E_{\rm rot}$ & $B_{\rm dip}$ & $B_{\rm quad}$\\ 
        &         &          & (erg) & (ms) & (erg) & (G) & (G) \\ 
        \hline
130427A & BdHN I  & 0.34     & $1.40\times10^{54}$& 0.95  & $3.50\times10^{52}$ & $6.0 \times 10^{12}$ & $2.0 \times 10^{13} \sim 6.0 \times 10^{14}$  \\
160509A & BdHN I  & 1.17     & $1.06\times10^{54}$& 0.75  & $5.61\times10^{52}$ & $4.0 \times 10^{12}$ & $1.3 \times 10^{14} \sim 4.0 \times 10^{14}$  \\
160625B & BdHN I  & 1.406    & $3.00\times10^{54}$& 0.5  & $1.26\times10^{53}$ & $1.5 \times 10^{12}$ & $5.0 \times 10^{13} \sim 1.6 \times 10^{14}$  \\
190114C & BdHN I  & 0.42     & $2.47\times10^{53}$& 2.1  & $7.16\times10^{51}$ & $5.0 \times 10^{12}$ & $1.5 \times 10^{15} \sim 5.0 \times 10^{15}$  \\\hline
180728A & BdHN II & 0.117    & $2.73\times10^{51}$& 3.5  & $2.58\times10^{51}$ & $1.0 \times 10^{13}$ & $3.5 \times 10^{15} \sim 1.1 \times 10^{16}$  \\

\hline
\end{tabular}
\end{table*}

\section{A self-consistency check}\label{sec:3.3}

Having estimated the magnetic field structure and the rotation period of the $\nu$NS from the fit of the data of the FPA phase at times $10^2$--$10^7$~s, we can now assess their self-consistency with expected values within the BdHN scenario.

First, let us adopt the binary as tidally locked, i.e. the rotation period of the binary components is synchronized with the orbital period. This implies that the rotation period of the CO$_{\rm core}$ is $P_{\rm CO} = P$, where $P$ denotes the orbital period. From the Kepler law the value of $P$ is connected to the orbital separation $a_{\rm orb}$ and with the binary mass as:
\begin{equation}
    P_{\rm CO} = P = 2 \pi \sqrt{\frac{a^3_{\rm orb}}{G M_{\rm tot}}},
\label{eq:CO_period}
\end{equation}
where $G$ is the gravitational constant and $M_{\rm tot}=M_{\rm CO}+M_{\rm NS}$ is the total mass of the binary, where $M_{\rm CO}$ and $M_{\rm NS}$ are the masses of the CO$_{\rm core}$ and the NS companion, respectively. Thus, $M_{\rm CO}=M_{\rm Fe} + M_{\rm ej}$ with $M_{\rm Fe}$ and $M_{\rm ej}$ the masses of the iron core (which collapses and forms the $\nu$NS) and the ejected mass in the SN event, respectively.

The mass of the $\nu$NS is $M_{\nu\rm NS}\approx M_{\rm Fe}$. The rotation period, $P_{\nu\rm NS}$, is estimated from the one of the iron core, $P_{\rm Fe}$, by applying the angular momentum conservation in the collapse process, i.e.:
\begin{equation}
    P_{\nu\rm NS} = \left(\frac{R_{\nu\rm NS}}{R_{\rm Fe}}\right)^2 P,
\label{eq:NS_period}
\end{equation}
where $R_{\nu\rm NS}$ and $R_{\rm Fe}$ are the radius of the $\nu$NS and of the iron core, respectively, and we have assumed that the pre-SN star has uniform rotation; so $P_{\rm Fe} = P_{\rm CO} = P$.

Without loss of generality, in our estimates we can adopt a $\nu$NS order-of-magnitude radius of $10^6$~cm. As we shall see below, a more careful estimate is the one of the CO$_{\rm core}$ progenitor (which tell us the radius of the iron core) and the orbital period/binary separation which affect additional observables of a BdHN.

It is instructive to appreciate the above statement with specific examples; for which we use the results of \citet{2019ApJ...874...39W} for two BdHN archetypes: GRB 130427A for BdHN I and GRB 180827A for BdHN II. Table~\ref{tab:ns_parameters} shows, for the above GRBs, as well as for GRB 190114C, GRB 160625B and GRB 160509A, some observational quantities (the isotropic energy released $E_{\rm iso}$ and the cosmological redshift), the inferred BdHN type and the properties of the $\nu$NS (rotation period $P_{\nu \rm NS}$, rotational energy and the strength of the dipole and quadrupole magnetic field components).

By examining the BdHN models simulated in \citet{2019ApJ...871...14B} (see e.g. Table~2 there), we have shown in \citet{2019ApJ...874...39W} that the Model `25m1p08e' fits the observational requirements of GRB 130427A, and the Model `25m3p1e' the ones of GRB 180827A. These models have the same binary progenitor components: the $\approx 6.8~M_\odot$ CO$_{\rm core}$ ($R_{\rm Fe} \sim 2 \times 10^8$~cm) developed by a $25 M_{\odot}$ ZAMS star (see Table~1 in \citealp{2019ApJ...871...14B}) and a $2~M_{\odot}$ NS companion. For GRB 130427A the orbital period is $P=4.8$~min (binary separation $a_{\rm orb}\approx 1.3\times 10^{10}$~cm), resulting in $P_{\nu\rm NS} \approx 1.0$~ms while, for GRB 180827A, the orbital period is $P=11.8$~min ($a_{\rm orb}\approx 2.6\times 10^{10}$~cm) so a less compact binary, which leads to $P_{\nu\rm NS} \approx 2.5$~ms.

We turn now to perform a further self-consistency check of our picture. Namely, we make a cross-check of the estimated $\nu$NS parameters obtained first from the early afterglow via synchrotron emission, and then from the late X-ray afterglow via the pulsar luminosity, with respect to expectations from NS theory. 

Up to factors of order unity, the surface dipole $B_s$ and the toroidal component $B_t$ at a distance $r$ from the surface are approximately related as \citep[see, e.g.,][]{1969ApJ...157..869G})
\begin{equation}\label{eq:BtBs}
    B_t \approx \left( \frac{2 \pi R_{\nu\rm NS}}{c P_{\nu\rm NS}} \right)^2 \left( \frac{R_{\nu\rm NS}}{r} \right)B_s.
\end{equation}

Let us analyze the case of GRB 130427A. By equating  Eqs.~(\ref{eq:magneticField}) and (\ref{eq:BtBs}), and using the values of $B_0 = 5 \times 10^5$~Gauss and $R_0 = 2.4 \times 10^{12}$~cm from \citet{2018ApJ...869..101R} obtained from the synchrotron analysis, and $P_{\nu\rm NS} = P_0 \approx 1$~ms from the pulsar activity in the late afterglow analysis, we obtain $B_s \approx 2\times 10^{13}$~G. This value has to be compared with the one obtained from the request that the pulsar luminosity powers the late afterglow, $B_{\rm dip} = 6\times 10^{12}$~G (see Table~\ref{tab:ns_parameters}). If we use the parameters $B_0 = 1.0 \times 10^{6}$~Gauss and $R_0 = 1.2 \times 10^{11}$~cm from Table~\ref{tab:parameters2} for GRB 160625B, and the corresponding $P_{\nu\rm NS} = P_0 \approx 0.5$~ms, we obtain $B_s \approx 6.8\times 10^{11}$~G, to be compared with $B_{\rm dip} \approx 10^{12}$~G (see Table~\ref{tab:ns_parameters}). An even better agreement can be obtained by using a more accurate value of the $\nu$NS radius which is surely bigger than the fiducial value $R_{\nu\rm NS}= 10^6$~cm we have used in these estimates.

\section{Nature of the dipole+quadrupole magnetic field structure of the $\nu$NS}\label{sec:5}

We attribute the spin-down energy of the $\nu$NS to the energy injection of the late-time afterglow. By fitting the observed emission through the synchrotron model, the spin period and the magnetic field of the $\nu$NS can be inferred. In \citet{2019ApJ...874...39W}, we have applied this approach on GRB 130427A and GRB 180728A, here we apply the same method on the recent GRB 190114C and other two, GRB 160509A and GRB 160625B, for comparison. As shown in Fig.~\ref{fig:lightcurves}, we plot the energy injection from the dipole and quadruple emission of $\nu$NS, the fitting results indicate 190114C leaves a $\nu$NS of spin period $2.1$~ms, with dipole magnetic field $B_{dip} = 5 \times 10^{12}$~G, and a quadruple magnetic field $>10^{15}$~G, the fitting parameters of all the GRBs are listed in table \ref{tab:ns_parameters}. Generally, the NS in the BdHN I system spins faster, of period $\lesssim 2$~ms, and contains more rotational energy $\gtrsim 10^{52}$~erg. We notice that GRB 160625B has the shortest initial spin period of $P = 0.5$~ms, which is exactly on the margin of the rotational period of a NS at the Keplerian sequence. For a NS of mass $1.4~M_{\odot}$ and radius $12$~km, its Keplerian frequence $f_{\rm K} \simeq 1900$ \citep{2004Sci...304..536L,2019PhRvD..99d3004R}, corresponding to the spin period of $P \simeq 0.5$~ms. 

From Eq.~(\ref{eq:CO_period}) and (\ref{eq:NS_period}), the orbital separation of binary system relates to the spin of $\nu$NS, $a_{\rm orb} \propto  P_{\nu\rm NS}^{2/3}$. Therefore, with the knowledge of the binary separation of GRB 130427A $\sim 1.35 \times 10^{10}$~cm, the spin period of $\sim 1$~ms, and the newly inferred spin of GRB 190114C $\sim 1.2$~ms, assuming these two systems have the same mass and radius of the CO$_{\rm core}$ and the $\nu$NS, we obtain the orbital separation of GRB 190114C as $\sim 1.52 \times 10^{10}$~cm.

The self-consistent value obtained for the orbital period/separation give a strong support to our basic assumptions: 1) owing to the system compactness the binary components are tidally locked, and 2) angular momentum is conserved in the core-collapse SN process.

We would like to recall that it has been shown that purely poloidal field configurations are unstable against adiabatic perturbations; for non-rotating stars it has been first demonstrated by \citet{1973MNRAS.162..339W,1973MNRAS.163...77M}; see also \citet{1977ApJ...215..302F}. For rotating stars similar results have been obtained, e.g., by~\citet{1985MNRAS.216..139P}. In addition, \citet{1973MNRAS.161..365T} has shown that purely toroidal configurations are also unstable. We refer the reader to \citet{1999A&A...349..189S} for a review on the different possible instabilities that may be active in magnetic stars. In this line, the dipole-quadrupole magnetic field configuration found in our analyses with a quadrupole component dominating in the early life of the the $\nu$NS are particularly relevant. They also give support to theoretical expectations pointing to the possible stability of poloidal-toroidal magnetic field configurations on timescales longer than the collapsing time of the pre-SN star; see e.g. for details \citet{1980MNRAS.191..151T,1984AN....305..301M}.

It remains the question of how, during the process of gravitational collapse, the magnetic field increase its strength to the NS observed values. This is still one of the most relevant open questions in astrophysics which is at this stage out of the scope of the present work. We shall mention here only one important case which is the traditional explanation of the NS magnetic field strength based on the amplification of the field by magnetic flux conservation. The flux conservation implies $\Phi_i = \pi B_i R_i^2 = \Phi_f = \pi B_f R_f^2$, where $i$ and $f$ stand for initial and final configurations and $R_{i,f}$ the corresponding radii. The radius of the collapsing iron core is of the order of $10^8$--$10^9$~cm while the radius of the $\nu$NS is of the order of $10^6$~cm; therefore, the magnetic flux conservation implies an amplification of $10^4$--$10^6$ times the initial field during the $\nu$NS formation. Therefore, a seed magnetic field of $10^7$--$10^9$~G is necessary to be present in the iron core of the pre-SN star to explain a $\nu$NS magnetic field of $10^{13}$~G. The highest magnetic fields observed in main-sequence stars leading to the pre-SN stars of interest are of the order of $10^4$~G \citep{2009IAUS..259...61S}. If the magnetic field is uniform inside the star, then the value of the magnetic field observed in these stars poses a serious issue to the magnetic flux conservation hypothesis for the NS magnetic field genesis. A summary of the theoretical efforts to understand the possible sources of the magnetic field of a NS can be found in \citet{2009IAUS..259...61S}. 

\section{Nature of the magnetic field around the newborn BH}\label{sec:6}

The BH in a BdHN I is formed from the gravitational collapse of the NS companion of the CO$_{\rm core}$, which reaches the critical mass by the hypercritical accretion of the ejecta of the SN explosion of the CO$_{\rm core}$. Hence, the magnetic field surrounding the BH derived in the previous section for the explanation of the GeV emission should originate from the collapsed NS. In fact, the magnetic field of the $\nu$NS evaluated at the BH position is too low to be relevant in this discussion. As we shall see, the magnetic field inherited from the collapsed NS can easily reach values of the order of $10^{14}$~G. Instead, the magnetic field of the $\nu$NS at the BH site is $B_{\rm dip}\,(R_{\rm \nu NS}/a_{\rm orb})^3=10$~G, adopting fiducial parameters according to the results of Table~\ref{tab:ns_parameters}: a dipole magnetic field at the $\nu$NS surface $B_{\rm dip}=10^{13}$~G, a binary separation of $a_{\rm orb}= 10^{10}$~cm and a $\nu$NS radius of $R_{\rm \nu NS}=10^6$~cm.

{Having clarified this issue, we proceed now to discuss the nature of the field. Both the $\nu$NS and the NS follow an analogous formation channel, namely they are born from core-collapse SNe}. In fact, to reach the BdHN stage the massive binary has to survive two SN events: the first SN which forms the NS and the second one which forms the $\nu$NS (core-collapse of the CO$_{\rm core}$). Figure~\ref{fig:BdHN} shows the evolutionary path of a massive binary leading to a BdHN I. It is then clear that the NS companion of the CO$_{\rm core}$ will have magnetic field properties analogous to the ones of the $\nu$NS, and discussed in the previous section. {Therefore, we can conclude that the BH forms from the collapse of a magnetized and fast rotating NS.}

In this scenario, the magnetic field of the collapsing NS companion should then be responsible of the magnetic field surrounding the BH. It is needed only a modest amplification of the initial field from the NS, which is $\sim 10^{13}$~G, to reach the value of $10^{14}$~G around the newborn BH. Then, even the single action of magnetic flux conservation can suffice to explain the magnetic field amplification. The BH horizon is $r_+\sim G M/c^2$, where $M$ can be assumed to be equal to the NS critical mass, say $3~M_\odot$, so $r_+\approx 4.4$~km. The NS at the collapse point, owing to high rotation, will have a radius in excess of the typically adopted $10$~km \citep{2015PhRvD..92b3007C}; let us assume a conservative range $12$--$15$~km. These conditions suggest that magnetic flux conservation magnifies the magnetic field in the BH formation by a factor $7$--$12$. Therefore, a seed field of $10^{13}$~G present in the collapsing NS is enough to explain the magnetic field of $10^{14}$~G near the newborn BH.

\begin{figure*}
    \centering
    (a)
    \includegraphics[width=0.44\hsize,clip]{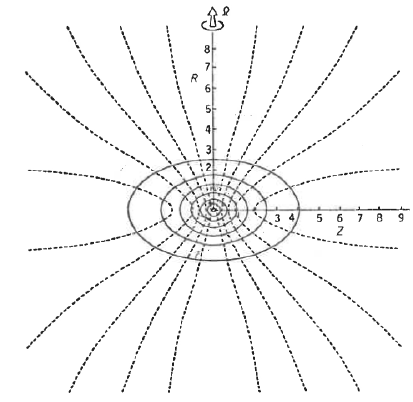}
    (b)
    \includegraphics[width=0.45\hsize,clip]{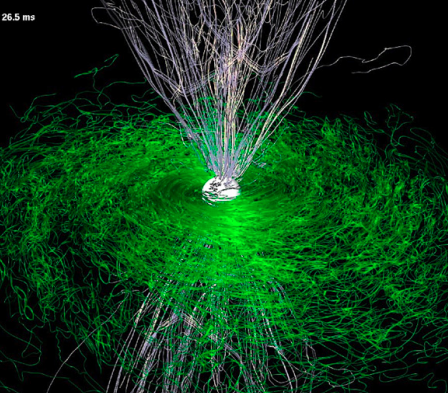}
    \caption{(a) Figure reproduced from \citet{1978pans.proc..644W}: numerical simulation of the gravitational collapse of a star accounting for the magnetic field presence. Isodensity surfaces are indicated by the solid lines and poloidal field lines are indicated by the dashed lines. The time is the end of the numerical simulation. (b) Figure taken from \citet{2011ApJ...732L...6R} by author's permission. Magnetic-field structure after the collapse to BH. Green refers to magnetic-field lines inside the torus and on the equatorial plane, while white refers to magnetic-field lines outside the torus and near the axis.}
    \label{fig:wilson}
\end{figure*}

It is worthy to clarify a crucial point: the magnetic field has to remain anchored to some NS material which guarantee its existence. It is therefore expected that some part of the NS does not take part of the BH formation. Assuming that magnetic flux is conserved during the collapse, then the magnetic energy is a constant fraction of the gravitational energy during the entire process, so only high rotation \citep[see, e.g.,][]{2016ApJ...833..107B} and some degree of differential rotation \citep[see, e.g.,][]{2006PhRvL..96c1102S} of the NS at the critical mass point can be the responsible of avoiding some fraction of NS matter to remain outside with sufficient angular momentum to orbit the newborn BH {(see, e.g., Fig.~\ref{fig:wilson})}.

The three-dimensional simulations of BdHNe presented in \citet{2019ApJ...871...14B} show that the part of the SN ejecta surrounding the BH forms a torus-like structure around it. The aforementioned matter from the NS with high angular momentum will add to this orbiting matter around the BH. In the off-equatorial directions the density is much smaller \citep[see also]{2018ApJ...869..101R,2019ApJ...871...14B,2019arXiv190404162R}. This implies that on the equatorial plane the field is compressed while in the axial direction the matter accretion flows in along the field lines.

Our \emph{inner engine}, the BH$+$magnetic field configuration powering the high-energy emission in a BdHN I finds additional support in numerical simulations of magnetic {and rotating} collapse into a BH. The first numerical computer treatment of the gravitational collapse to a BH in presence of magnetic fields, starts with the pioneering two-dimensional simulations by \citet{1975NYASA.262..123W} (see Fig.~\ref{fig:wilson} (a) reproduced from \citealp{1978pans.proc..644W}). These works already showed the amplification of the magnetic field in the gravitational collapse process. Rotating magnetized gravitational collapse into a BH has been more recently treated with greater detail by three-dimensional simulations which have confirmed this picture and the crucial role of the combined presence of magnetic field and rotation \citep{2013PhRvD..88d4020D,2017MNRAS.469L..31N,2018ApJ...864..117M}.

{
Additional support can be also found in the context of the binary NS mergers. Numerical simulations have indeed shown that the collapse of the unstable massive NS formed in the merger into a BH leads to a configuration composed of a BH surrounded by a nearly collimated magnetic field and an accretion disk
\citep[see][for details]{2006PhRvL..96c1101D,2006PhRvL..96c1102S,2006PhRvD..73j4015D,2007CQGra..24S.207S,2008PhRvD..77d4001S}. Three-dimensional numerical simulation have been also performed and confirm this scenario \citep{2011ApJ...732L...6R}. In particular, it is appropriate to underline the strong analogy between Fig.~\ref{fig:wilson} (a) taken from \citet{1978pans.proc..644W} with Fig.~\ref{fig:wilson} (b) reproduced in this paper from \citet{2011ApJ...732L...6R}}. It is also interesting the value of the magnetic field close to the BH estimated in \citet{2011ApJ...732L...6R}, along the BH spin axis, $8 \times 10^{14}$~G , similar to the value of $3\times 10^{14}$~G needed for the operation of the ``\textit{inner engine}'' of GRB 130427A \citep{2018arXiv181101839R}. What is also conceptually important is that the uniform magnetic field assumed by the Wald solution should be expected to reach a poloidal configuration already relatively close to the BH. This occurs already in the original \citet{1978pans.proc..644W} solution confirmed by the recent and most detailed calculation by \citet{2011ApJ...732L...6R}, see Fig.~\ref{fig:wilson} (a) and (b).

Although the above simulations refer to the remnant configuration of a binary NS merger, the post-merger configuration is analogous to the one developed in BdHNe I related to the newborn BH, which we have applied in our recent works \citep[see e.g.][and references therein]{2018ApJ...869..101R,2018arXiv181101839R,2019ApJ...886...82R,2019arXiv190404162R,2019ApJ...874...39W}, and which is supported by the recently presented three-dimensional simulations of BdHNe \citep[see][for details]{2019ApJ...871...14B}.

\begin{figure*}
\centering
\includegraphics[width=0.49\hsize,clip]{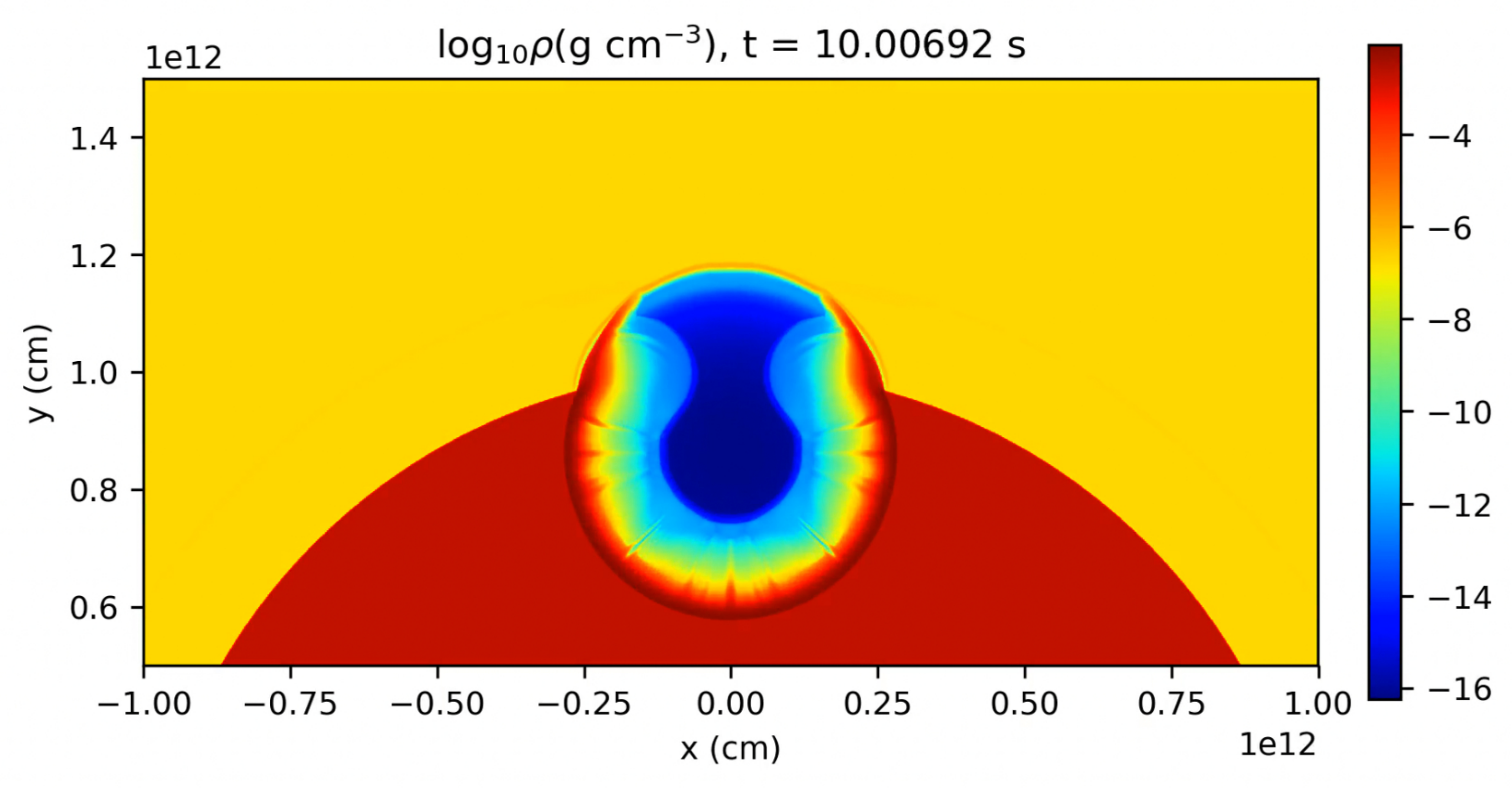}
\includegraphics[width=0.49\hsize,clip]{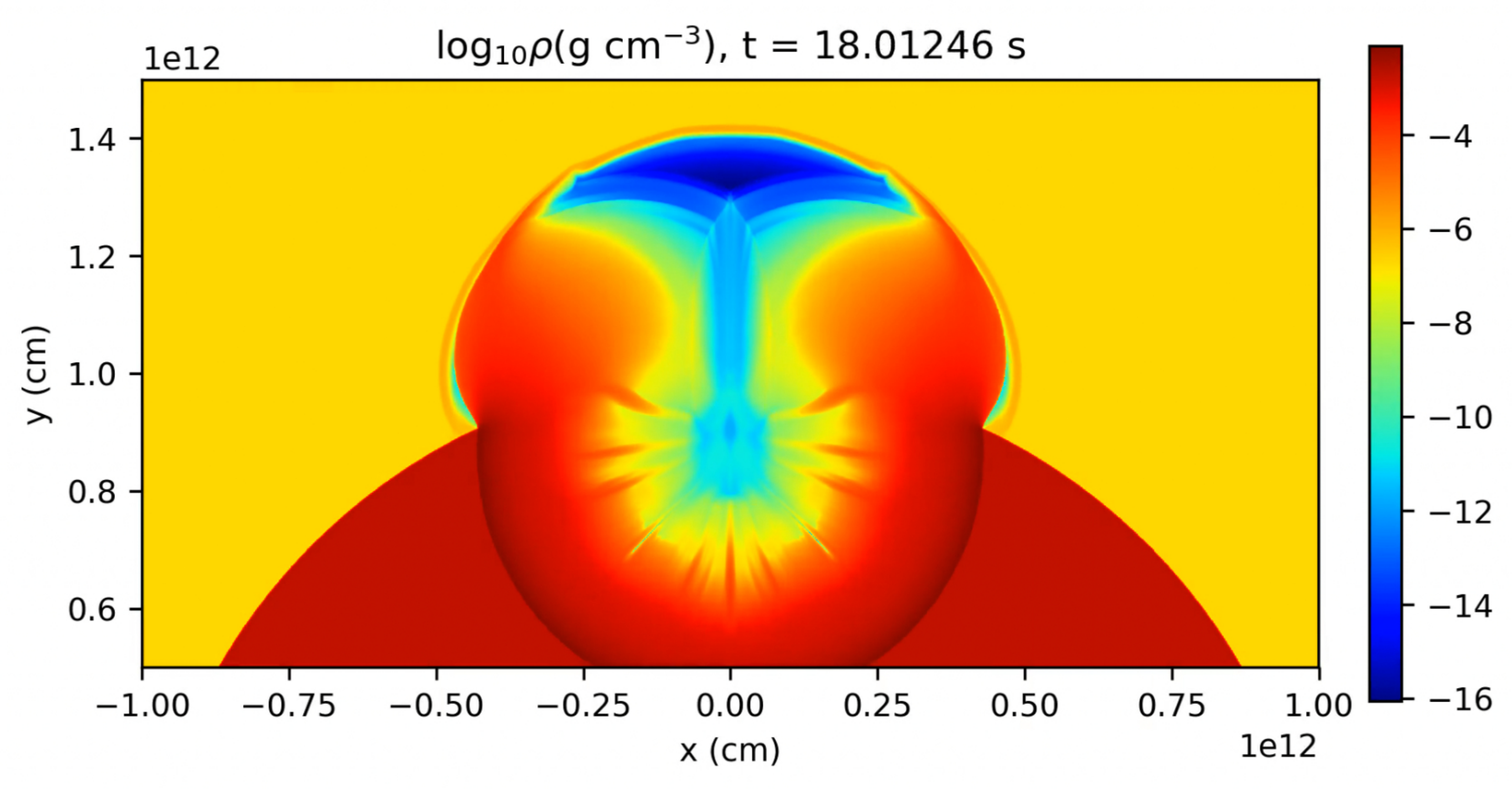}
\caption{{Spatial distribution of matter density at different time of impact of the $e^+~e^-~\gamma$ plasma onto the cavity walls at $t_{\rm imp} = 10$~s (left) and $t_{\rm imp} = 18$~s (right) for GRB 190114C; more information in \citet{2019ApJ...883..191R} }.}
\label{cavity}
\end{figure*}

Before closing, let us indicate the difference between the NS merger and the BdHN. In the case of BdHN the gravitational collapse leading to the BH with the formation of an horizon creates a very-low-density cavity of $10^{-14}$~g~cm$^{-3}$ with a radius $\sim10^{11}$~cm in the SN ejecta, see Fig.~\ref{fig:BdHN} and Fig.~\ref{cavity}, reproduced from \citet{2019ApJ...883..191R}. The presence of such low-density environment is indeed essential for the successful operation of the ``\textit{inner engine}''.

Both the reaching of a poloidal configuration already close to the BH in the Wald solution, as well as the existing of the cavity are crucial factors in the analysis of the propagation of the photons produced by synchrotron radiation and the fulfilment of reaching the transparency condition by the ``\textit{inner engine}'' of the BdHNe \citep{2019ApJ...886...82R}.

\section{Conclusions}\label{sec:7}
Our general conclusions have been reached based on the comparison and contrast of the observations of GRB 130427A, GRB 160509A, GRB 160625B, GRB 180728A and GRB 190114C:

 
\begin{enumerate}
    \item
    From the analysis of GRB 130427A \citep{2018ApJ...869..151R} and GRB 190114C presented here (see Figs.~\ref{fig:lum_comparison} and \ref{fig:lightcurves}), we conclude that the early ($t\sim 10^2$--$10^4$~s) X-ray emission of the FPA phase is explained by the injection of ultra-relativistic electrons from the $\nu$NS into the magnetized expanding ejecta, producing synchrotron radiation. The magnetic field inferred in this part of the analysis is found to be consistent with the toroidal/longitudinal magnetic field component of the $\nu$NS. The dominance of this component is expected at distances much larger ($\sim 10^{12}$~cm) than the light cylinder radius in which this synchrotron emission occurs. No data of the other GRBs considred in this paper are available in this time interval.
    \item
    Using the data of all the present GRBs, we concluded that at times $t\gtrsim 10^3$--$10^4$~s of the FPA phase, the power-law decaying luminosity is dominated via the pulsar magnetic-braking radiation. We have inferred a dipole+quadrupole structure of the $\nu$NS magnetic field, being the quadrupole component initially dominant. The strength of the dipole component is about $10^{12}$--$10^{13}$~G while the one of the quadrupole can be of order $10^{15}$~G (see Fig.~\ref{fig:lum_comparison} and Table~\ref{tab:ns_parameters}). As clearly shown in Figs.~\ref{fig:lum_comparison} and \ref{fig:lightcurves}, the solely $\nu$NS with the dipole+quadrupole magnetic field structure can not explain the emission of the early FPA phase which is dominated by the SN emission.
    \item
    We have checked that the magnetic field of the $\nu$NS, inferred independently in the two above regimes of the FPA phase, give values in very good agreement. The $\nu$NS magnetic field obtained from the explanation of the FPA phase, at times $10^2$--$10^3$~s by synchrotron radiation, and at times $t\gtrsim 10^4$~s by pulsar magnetic-braking, are in close agreement (see Sec.~\ref{sec:4}, Table~\ref{tab:ns_parameters} and Fig.~\ref{fig:lightcurves}).
    \item
    In section~\ref{sec:3.3}, we have shown the consistency of the inferred $\nu$NS parameters with the expectations in the BdHN scenario. In particular, we have used the rotation period of the $\nu$NS inferred from the FPA phase at times $t\gtrsim 10^3$--$10^4$~s, we have inferred the orbital period/separation assuming tidal synchronization of the binary and angular momentum conservation in the gravitational collapse of the iron core leading to the $\nu$NS. This inferred binary separation is shown to be in excellent agreement with the numerical simulations of the binary progenitor in \citet{2019ApJ...874...39W}.

\end{enumerate}

Before concluding, in view of the recent understanding gained on the ``\textit{inner engine}'' of the high-energy emission of the GRB \citep{2019ApJ...886...82R}, we can also conclude:

\begin{enumerate}
    \item[5.]
    The magnetic field along the rotational axis of the BH is rooted in the magnetosphere left by the binary companion NS prior to the collapse.
    \item[6.]
    While in the equatorial plane the field is magnified by magnetic flux conservation, in the axial direction the matter accretion flows in along the field lines; see Fig.~\ref{fig:25Mzams2Mns} and \citet{2019ApJ...871...14B}. Indeed, three-dimensional numerical simulations of the gravitational collapse into a BH in presence of rotation and magnetic field confirm our picture; see Fig.~\ref{fig:wilson} and \citet{2011ApJ...732L...6R,2013PhRvD..88d4020D,2017MNRAS.469L..31N,2018ApJ...864..117M}.
    \item[7.]
    The clarification reached in the role of the SN accretion both on the NS and on the $\nu$NS, the stringent limits imposed on the Lorentz factor of the FPA phase, the energetic requirement of the ``\textit{inner engine}'' inferred from the recent publications, clearly points to an electrodynamical nature of the ``\textit{inner engine}'' of the GRB, occurring close to the BH horizon, as opposed to the traditional, gravitational massive blastwave model.

\end{enumerate}

\begin{acknowledgements}
We acknowledge the public data from Swift and Fermi satellites. We appreciate the discussion with Prof. She-sheng Xue, and the suggestions from the referee.
\end{acknowledgements}

\bibliographystyle{aasjournal}
\bibliography{biblio}

\begin{thebibliography}{}
\expandafter\ifx\csname natexlab\endcsname\relax\def\natexlab#1{#1}\fi
\providecommand{\url}[1]{\href{#1}{#1}}
\providecommand{\dodoi}[1]{doi:~\href{http://doi.org/#1}{\nolinkurl{#1}}}
\providecommand{\doeprint}[1]{\href{http://ascl.net/#1}{\nolinkurl{http://ascl.net/#1}}}
\providecommand{\doarXiv}[1]{\href{https://arxiv.org/abs/#1}{\nolinkurl{https://arxiv.org/abs/#1}}}

\bibitem[{{Ackermann} {et~al.}(2014){Ackermann}, {Ajello}, {Asano}, {Atwood},
  {Axelsson}, {Baldini}, {Ballet}, {Barbiellini}, {Baring}, {Bastieri},
  {Bechtol}, {Bellazzini}, {Bissaldi}, {Bonamente}, {Bregeon}, {Brigida},
  {Bruel}, {Buehler}, {Burgess}, {Buson}, {Caliandro}, {Cameron}, {Caraveo},
  {Cecchi}, {Chaplin}, {Charles}, {Chekhtman}, {Cheung}, {Chiang}, {Chiaro},
  {Ciprini}, {Claus}, {Cleveland}, {Cohen-Tanugi}, {Collazzi}, {Cominsky},
  {Connaughton}, {Conrad}, {Cutini}, {D'Ammando}, {de Angelis}, {DeKlotz}, {de
  Palma}, {Dermer}, {Desiante}, {Diekmann}, {Di Venere}, {Drell},
  {Drlica-Wagner}, {Favuzzi}, {Fegan}, {Ferrara}, {Finke}, {Fitzpatrick},
  {Focke}, {Franckowiak}, {Fukazawa}, {Funk}, {Fusco}, {Gargano}, {Gehrels},
  {Germani}, {Gibby}, {Giglietto}, {Giles}, {Giordano}, {Giroletti}, {Godfrey},
  {Granot}, {Grenier}, {Grove}, {Gruber}, {Guiriec}, {Hadasch}, {Hanabata},
  {Harding}, {Hayashida}, {Hays}, {Horan}, {Hughes}, {Inoue}, {Jogler},
  {J{\'o}hannesson}, {Johnson}, {Kawano}, {Kn{\"o}dlseder}, {Kocevski}, {Kuss},
  {Lande}, {Larsson}, {Latronico}, {Longo}, {Loparco}, {Lovellette}, {Lubrano},
  {Mayer}, {Mazziotta}, {McEnery}, {Michelson}, {Mizuno}, {Moiseev}, {Monzani},
  {Moretti}, {Morselli}, {Moskalenko}, {Murgia}, {Nemmen}, {Nuss}, {Ohno},
  {Ohsugi}, {Okumura}, {Omodei}, {Orienti}, {Paneque}, {Pelassa}, {Perkins},
  {Pesce-Rollins}, {Petrosian}, {Piron}, {Pivato}, {Porter}, {Racusin},
  {Rain{\`o}}, {Rando}, {Razzano}, {Razzaque}, {Reimer}, {Reimer}, {Ritz},
  {Roth}, {Ryde}, {Sartori}, {Parkinson}, {Scargle}, {Schulz}, {Sgr{\`o}},
  {Siskind}, {Sonbas}, {Spandre}, {Spinelli}, {Tajima}, {Takahashi}, {Thayer},
  {Thayer}, {Thompson}, {Tibaldo}, {Tinivella}, {Torres}, {Tosti}, {Troja},
  {Usher}, {Vandenbroucke}, {Vasileiou}, {Vianello}, {Vitale}, {Winer}, {Wood},
  {Yamazaki}, {Younes}, {Yu}, {Zhu}, {Bhat}, {Briggs}, {Byrne}, {Foley},
  {Goldstein}, {Jenke}, {Kippen}, {Kouveliotou}, {McBreen}, {Meegan},
  {Paciesas}, {Preece}, {Rau}, {Tierney}, {van der Horst}, {von Kienlin},
  {Wilson-Hodge}, {Xiong}, {Cusumano}, {La Parola}, \&
  {Cummings}}]{2014Sci...343...42A}
{Ackermann}, M., {Ajello}, M., {Asano}, K., {et~al.} 2014, Science, 343, 42,
  \dodoi{10.1126/science.1242353}

\bibitem[{{Alexander} {et~al.}(2017){Alexander}, {Laskar}, {Berger},
  {Guidorzi}, {Dichiara}, {Fong}, {Gomboc}, {Kobayashi}, {Kopac}, {Mundell},
  {Tanvir}, \& {Williams}}]{2017ApJ...848...69A}
{Alexander}, K.~D., {Laskar}, T., {Berger}, E., {et~al.} 2017, \apj, 848, 69,
  \dodoi{10.3847/1538-4357/aa8a76}

\bibitem[{{Becerra} {et~al.}(2016){Becerra}, {Bianco}, {Fryer}, {Rueda}, \&
  {Ruffini}}]{2016ApJ...833..107B}
{Becerra}, L., {Bianco}, C.~L., {Fryer}, C.~L., {Rueda}, J.~A., \& {Ruffini},
  R. 2016, \apj, 833, 107, \dodoi{10.3847/1538-4357/833/1/107}

\bibitem[{{Becerra} {et~al.}(2015){Becerra}, {Cipolletta}, {Fryer}, {Rueda}, \&
  {Ruffini}}]{2015ApJ...812..100B}
{Becerra}, L., {Cipolletta}, F., {Fryer}, C.~L., {Rueda}, J.~A., \& {Ruffini},
  R. 2015, \apj, 812, 100, \dodoi{10.1088/0004-637X/812/2/100}

\bibitem[{{Becerra} {et~al.}(2019){Becerra}, {Ellinger}, {Fryer}, {Rueda}, \&
  {Ruffini}}]{2019ApJ...871...14B}
{Becerra}, L., {Ellinger}, C.~L., {Fryer}, C.~L., {Rueda}, J.~A., \& {Ruffini},
  R. 2019, \apj, 871, 14, \dodoi{10.3847/1538-4357/aaf6b3}

\bibitem[{{Becerra} {et~al.}(2018){Becerra}, {Guzzo}, {Rossi-Torres}, {Rueda},
  {Ruffini}, \& {Uribe}}]{2018ApJ...852..120B}
{Becerra}, L., {Guzzo}, M.~M., {Rossi-Torres}, F., {et~al.} 2018, \apj, 852,
  120, \dodoi{10.3847/1538-4357/aaa296}

\bibitem[{{Bianco} {et~al.}(2001){Bianco}, {Ruffini}, \&
  {Xue}}]{2001A&A...368..377B}
{Bianco}, C.~L., {Ruffini}, R., \& {Xue}, S.-S. 2001, \aap, 368, 377,
  \dodoi{10.1051/0004-6361:20000556}

\bibitem[{{Blandford} \& {McKee}(1976)}]{1976PhFl...19.1130B}
{Blandford}, R.~D., \& {McKee}, C.~F. 1976, Physics of Fluids, 19, 1130,
  \dodoi{10.1063/1.861619}

\bibitem[{{Blandford} \& {Znajek}(1977)}]{1977MNRAS.179..433B}
{Blandford}, R.~D., \& {Znajek}, R.~L. 1977, \mnras, 179, 433,
  \dodoi{10.1093/mnras/179.3.433}

\bibitem[{{Cipolletta} {et~al.}(2015){Cipolletta}, {Cherubini}, {Filippi},
  {Rueda}, \& {Ruffini}}]{2015PhRvD..92b3007C}
{Cipolletta}, F., {Cherubini}, C., {Filippi}, S., {Rueda}, J.~A., \& {Ruffini},
  R. 2015, \prd, 92, 023007, \dodoi{10.1103/PhysRevD.92.023007}

\bibitem[{{Costa} {et~al.}(1997){Costa}, {Frontera}, {Heise}, {Feroci}, {in't
  Zand}, {Fiore}, {Cinti}, {Dal Fiume}, {Nicastro}, {Orlandini}, {Palazzi},
  {Rapisarda}, {Zavattini}, {Jager}, {Parmar}, {Owens}, {Molendi}, {Cusumano},
  {Maccarone}, {Giarrusso}, {Coletta}, {Antonelli}, {Giommi}, {Muller}, {Piro},
  \& {Butler}}]{1997Natur.387..783C}
{Costa}, E., {Frontera}, F., {Heise}, J., {et~al.} 1997, \nat, 387, 783,
  \dodoi{10.1038/42885}

\bibitem[{{Dai} \& {Lu}(1998{\natexlab{a}})}]{1998A&A...333L..87D}
{Dai}, Z.~G., \& {Lu}, T. 1998{\natexlab{a}}, \aap, 333, L87

\bibitem[{{Dai} \& {Lu}(1998{\natexlab{b}})}]{1998PhRvL..81.4301D}
---. 1998{\natexlab{b}}, Physical Review Letters, 81, 4301,
  \dodoi{10.1103/PhysRevLett.81.4301}

\bibitem[{{Damour} \& {Ruffini}(1975)}]{1975PhRvL..35..463D}
{Damour}, T., \& {Ruffini}, R. 1975, Physical Review Letters, 35, 463,
  \dodoi{10.1103/PhysRevLett.35.463}

\bibitem[{{de Pasquale} {et~al.}(2007){de Pasquale}, {Oates}, {Page},
  {Burrows}, {Blustin}, {Zane}, {Mason}, {Roming}, {Palmer}, {Gehrels}, \&
  {Zhang}}]{2007MNRAS.377.1638D}
{de Pasquale}, M., {Oates}, S.~R., {Page}, M.~J., {et~al.} 2007, \mnras, 377,
  1638, \dodoi{10.1111/j.1365-2966.2007.11724.x}

\bibitem[{{Dionysopoulou} {et~al.}(2013){Dionysopoulou}, {Alic}, {Palenzuela},
  {Rezzolla}, \& {Giacomazzo}}]{2013PhRvD..88d4020D}
{Dionysopoulou}, K., {Alic}, D., {Palenzuela}, C., {Rezzolla}, L., \&
  {Giacomazzo}, B. 2013, \prd, 88, 044020, \dodoi{10.1103/PhysRevD.88.044020}

\bibitem[{{Duez} {et~al.}(2006{\natexlab{a}}){Duez}, {Liu}, {Shapiro},
  {Shibata}, \& {Stephens}}]{2006PhRvL..96c1101D}
{Duez}, M.~D., {Liu}, Y.~T., {Shapiro}, S.~L., {Shibata}, M., \& {Stephens},
  B.~C. 2006{\natexlab{a}}, Physical Review Letters, 96, 031101,
  \dodoi{10.1103/PhysRevLett.96.031101}

\bibitem[{{Duez} {et~al.}(2006{\natexlab{b}}){Duez}, {Liu}, {Shapiro},
  {Shibata}, \& {Stephens}}]{2006PhRvD..73j4015D}
---. 2006{\natexlab{b}}, \prd, 73, 104015, \dodoi{10.1103/PhysRevD.73.104015}

\bibitem[{{Fan} \& {Xu}(2006)}]{2006MNRAS.372L..19F}
{Fan}, Y.-Z., \& {Xu}, D. 2006, \mnras, 372, L19,
  \dodoi{10.1111/j.1745-3933.2006.00217.x}

\bibitem[{{Fan} {et~al.}(2013){Fan}, {Yu}, {Xu}, {Jin}, {Wu}, {Wei}, \&
  {Zhang}}]{2013ApJ...779L..25F}
{Fan}, Y.-Z., {Yu}, Y.-W., {Xu}, D., {et~al.} 2013, \apjl, 779, L25,
  \dodoi{10.1088/2041-8205/779/2/L25}

\bibitem[{{Filippenko} {et~al.}(1995){Filippenko}, {Barth}, {Matheson},
  {Armus}, {Brown}, {Espey}, {Fan}, {Goodrich}, {Ho}, {Junkkarinen}, {Koo},
  {Lehnert}, {Martel}, {Mazzarella}, {Miller}, {Smith}, {Tytler}, \&
  {Wirth}}]{1995ApJ...450L..11F}
{Filippenko}, A.~V., {Barth}, A.~J., {Matheson}, T., {et~al.} 1995, \apjl, 450,
  L11, \dodoi{10.1086/309659}

\bibitem[{{Flowers} \& {Ruderman}(1977)}]{1977ApJ...215..302F}
{Flowers}, E., \& {Ruderman}, M.~A. 1977, \apj, 215, 302,
  \dodoi{10.1086/155359}

\bibitem[{{Fryer} {et~al.}(2015){Fryer}, {Oliveira}, {Rueda}, \&
  {Ruffini}}]{2015PhRvL.115w1102F}
{Fryer}, C.~L., {Oliveira}, F.~G., {Rueda}, J.~A., \& {Ruffini}, R. 2015,
  Physical Review Letters, 115, 231102, \dodoi{10.1103/PhysRevLett.115.231102}

\bibitem[{{Fryer} {et~al.}(2014){Fryer}, {Rueda}, \&
  {Ruffini}}]{2014ApJ...793L..36F}
{Fryer}, C.~L., {Rueda}, J.~A., \& {Ruffini}, R. 2014, \apjl, 793, L36,
  \dodoi{10.1088/2041-8205/793/2/L36}

\bibitem[{Fryer {et~al.}(1999)Fryer, Woosley, \&
  Hartmann}]{1999ApJ...526..152F}
Fryer, C.~L., Woosley, S.~E., \& Hartmann, D.~H. 1999, The Astrophysical
  Journal, 526, 152

\bibitem[{{Galama} {et~al.}(1998){Galama}, {Vreeswijk}, {van Paradijs},
  {Kouveliotou}, {Augusteijn}, {B{\"o}hnhardt}, {Brewer}, {Doublier},
  {Gonzalez}, {Leibundgut}, {Lidman}, {Hainaut}, {Patat}, {Heise}, {in't Zand},
  {Hurley}, {Groot}, {Strom}, {Mazzali}, {Iwamoto}, {Nomoto}, {Umeda},
  {Nakamura}, {Young}, {Suzuki}, {Shigeyama}, {Koshut}, {Kippen}, {Robinson},
  {de Wildt}, {Wijers}, {Tanvir}, {Greiner}, {Pian}, {Palazzi}, {Frontera},
  {Masetti}, {Nicastro}, {Feroci}, {Costa}, {Piro}, {Peterson}, {Tinney},
  {Boyle}, {Cannon}, {Stathakis}, {Sadler}, {Begam}, \&
  {Ianna}}]{1998Natur.395..670G}
{Galama}, T.~J., {Vreeswijk}, P.~M., {van Paradijs}, J., {et~al.} 1998, \nat,
  395, 670, \dodoi{10.1038/27150}

\bibitem[{{Goldreich} \& {Julian}(1969)}]{1969ApJ...157..869G}
{Goldreich}, P., \& {Julian}, W.~H. 1969, \apj, 157, 869,
  \dodoi{10.1086/150119}

\bibitem[{{Heger} {et~al.}(2003){Heger}, {Fryer}, {Woosley}, {Langer}, \&
  {Hartmann}}]{2003ApJ...591..288H}
{Heger}, A., {Fryer}, C.~L., {Woosley}, S.~E., {Langer}, N., \& {Hartmann},
  D.~H. 2003, \apj, 591, 288, \dodoi{10.1086/375341}

\bibitem[{{Iwamoto} {et~al.}(2000){Iwamoto}, {Nakamura}, {Nomoto}, {Mazzali},
  {Danziger}, {Garnavich}, {Kirshner}, {Jha}, {Balam}, \&
  {Thorstensen}}]{2000ApJ...534..660I}
{Iwamoto}, K., {Nakamura}, T., {Nomoto}, K., {et~al.} 2000, \apj, 534, 660,
  \dodoi{10.1086/308761}

\bibitem[{{Izzo} {et~al.}(2019){Izzo}, {de Ugarte Postigo}, {Maeda},
  {Th{\"o}ne}, {Kann}, {Della Valle}, {Sagues Carracedo}, {Micha{\l}owski},
  {Schady}, {Schmidl}, {Selsing}, {Starling}, {Suzuki}, {Bensch}, {Bolmer},
  {Campana}, {Cano}, {Covino}, {Fynbo}, {Hartmann}, {Heintz}, {Hjorth},
  {Japelj}, {Kami{\'n}ski}, {Kaper}, {Kouveliotou}, {Kru{\.Z}y{\'n}ski},
  {Kwiatkowski}, {Leloudas}, {Levan}, {Malesani}, {Micha{\l}owski},
  {Piranomonte}, {Pugliese}, {Rossi}, {S{\'a}nchez-Ram{\'\i}rez}, {Schulze},
  {Steeghs}, {Tanvir}, {Ulaczyk}, {Vergani}, \&
  {Wiersema}}]{2019Natur.565..324I}
{Izzo}, L., {de Ugarte Postigo}, A., {Maeda}, K., {et~al.} 2019, \nat, 565,
  324, \dodoi{10.1038/s41586-018-0826-3}

\bibitem[{{Kouveliotou} {et~al.}(2013){Kouveliotou}, {Granot}, {Racusin},
  {Bellm}, {Vianello}, {Oates}, {Fryer}, {Boggs}, {Christensen}, {Craig},
  {Dermer}, {Gehrels}, {Hailey}, {Harrison}, {Melandri}, {McEnery}, {Mundell},
  {Stern}, {Tagliaferri}, \& {Zhang}}]{2013ApJ...779L...1K}
{Kouveliotou}, C., {Granot}, J., {Racusin}, J.~L., {et~al.} 2013, \apjl, 779,
  L1, \dodoi{10.1088/2041-8205/779/1/L1}

\bibitem[{{Laskar} {et~al.}(2016){Laskar}, {Alexander}, {Berger}, {Fong},
  {Margutti}, {Shivvers}, {Williams}, {Kopa{\v c}}, {Kobayashi}, {Mundell},
  {Gomboc}, {Zheng}, {Menten}, {Graham}, \& {Filippenko}}]{2016ApJ...833...88L}
{Laskar}, T., {Alexander}, K.~D., {Berger}, E., {et~al.} 2016, \apj, 833, 88,
  \dodoi{10.3847/1538-4357/833/1/88}

\bibitem[{{Lattimer} \& {Prakash}(2004)}]{2004Sci...304..536L}
{Lattimer}, J.~M., \& {Prakash}, M. 2004, Science, 304, 536,
  \dodoi{10.1126/science.1090720}

\bibitem[{{Levan} {et~al.}(2013){Levan}, {Cenko}, {Perley}, \&
  {Tanvir}}]{2013GCN..14455...1L}
{Levan}, A.~J., {Cenko}, S.~B., {Perley}, D.~A., \& {Tanvir}, N.~R. 2013, GCN
  Circ., 14455

\bibitem[{{Li}(2019)}]{2019ApJS..242...16L}
{Li}, L. 2019, \apjs, 242, 16, \dodoi{10.3847/1538-4365/ab1b78}

\bibitem[{{Li} {et~al.}(2018{\natexlab{a}}){Li}, {Wang}, {Shao}, {Wu}, {Huang},
  {Zhang}, {Ryde}, \& {Yu}}]{2018ApJS..234...26L}
{Li}, L., {Wang}, Y., {Shao}, L., {et~al.} 2018{\natexlab{a}}, \apjs, 234, 26,
  \dodoi{10.3847/1538-4365/aaa02a}

\bibitem[{{Li} {et~al.}(2018{\natexlab{b}}){Li}, {Wu}, {Lei}, {Dai}, {Liang},
  \& {Ryde}}]{2018ApJS..236...26L}
{Li}, L., {Wu}, X.-F., {Lei}, W.-H., {et~al.} 2018{\natexlab{b}}, \apjs, 236,
  26, \dodoi{10.3847/1538-4365/aabaf3}

\bibitem[{{Li} {et~al.}(2015){Li}, {Wu}, {Huang}, {Wang}, {Tang}, {Liang},
  {Zhang}, {Wang}, {Geng}, {Liang}, {Wei}, {Zhang}, \&
  {Ryde}}]{2015ApJ...805...13L}
{Li}, L., {Wu}, X.-F., {Huang}, Y.-F., {et~al.} 2015, \apj, 805, 13,
  \dodoi{10.1088/0004-637X/805/1/13}

\bibitem[{{L{\"u}} \& {Zhang}(2014)}]{2014ApJ...785...74L}
{L{\"u}}, H.-J., \& {Zhang}, B. 2014, \apj, 785, 74,
  \dodoi{10.1088/0004-637X/785/1/74}

\bibitem[{{L{\"u}} {et~al.}(2015){L{\"u}}, {Zhang}, {Lei}, {Li}, \&
  {Lasky}}]{2015ApJ...805...89L}
{L{\"u}}, H.-J., {Zhang}, B., {Lei}, W.-H., {Li}, Y., \& {Lasky}, P.~D. 2015,
  \apj, 805, 89, \dodoi{10.1088/0004-637X/805/2/89}

\bibitem[{{L{\"u}} {et~al.}(2017){L{\"u}}, {L{\"u}}, {Zhong}, {Huang}, {Zhang},
  {Lan}, {Xie}, {Lu}, \& {Liang}}]{2017ApJ...849...71L}
{L{\"u}}, H.-J., {L{\"u}}, J., {Zhong}, S.-Q., {et~al.} 2017, \apj, 849, 71,
  \dodoi{10.3847/1538-4357/aa8f99}

\bibitem[{{MAGIC Collaboration} {et~al.}(2019){MAGIC Collaboration}, {Acciari},
  {Ansoldi}, {Antonelli}, {Arbet Engels}, {Baack}, {Babi{\'c}}, {Banerjee},
  {Barres de Almeida}, {Barrio}, {Becerra Gonz{\'a}lez}, {Bednarek},
  {Bellizzi}, {Bernardini}, {Berti}, {Besenrieder}, {Bhattacharyya},
  {Bigongiari}, {Biland }, {Blanch}, {Bonnoli}, {Bo{\v{s}}njak}, {Busetto},
  {Carosi}, {Carosi}, {Ceribella}, {Chai}, {Chilingaryan}, {Cikota}, {Colak},
  {Colin}, {Colombo}, {Contreras}, {Cortina}, {Covino}, {D'Amico}, {D'Elia},
  {da Vela}, {Dazzi}, {de Angelis}, {de Lotto}, {Delfino}, {Delgado},
  {Depaoli}, {di Pierro}, {di Venere}, {Do Souto Espi{\~n}eira}, {Dominis
  Prester}, {Donini}, {Dorner}, {Doro}, {Elsaesser}, {Fallah Ramazani},
  {Fattorini}, {Fern{\'a}ndez-Barral}, {Ferrara}, {Fidalgo}, {Foffano},
  {Fonseca}, {Font}, {Fruck}, {Fukami}, {Gallozzi}, {Garc{\'\i}a L{\'o}pez},
  {Garczarczyk}, {Gasparyan}, {Gaug}, {Giglietto}, {Giordano}, {Godinovi{\'c}},
  {Green}, {Guberman}, {Hadasch}, {Hahn}, {Herrera}, {Hoang}, {Hrupec},
  {H{\"u}tten}, {Inada}, {Inoue}, {Ishio}, {Iwamura}, {Jouvin}, {Kerszberg},
  {Kubo}, {Kushida}, {Lamastra}, {Lelas}, {Leone}, {Lindfors}, {Lombardi},
  {Longo}, {L{\'o}pez}, {L{\'o}pez-Coto}, {L{\'o}pez-Oramas}, {Loporchio},
  {Machado de Oliveira Fraga}, {Maggio}, {Majumdar}, {Makariev}, {Mallamaci},
  {Maneva}, {Manganaro}, {Mannheim}, {Maraschi}, {Mariotti}, {Mart{\'\i}nez},
  {Masuda}, {Mazin}, {Mi{\'c}anovi{\'c}}, {Miceli}, {Minev}, {Miranda},
  {Mirzoyan}, {Molina}, {Moralejo}, {Morcuende}, {Moreno}, {Moretti},
  {Munar-Adrover}, {Neustroev}, {Nigro}, {Nilsson}, {Ninci}, {Nishijima},
  {Noda}, {Nogu{\'e}s}, {N{\"o}the}, {Nozaki}, {Paiano}, {Palacio},
  {Palatiello}, {Paneque}, {Paoletti}, {Paredes}, {Pe{\~n}il}, {Peresano},
  {Persic}, {Prada Moroni}, {Prand ini}, {Puljak}, {Rhode}, {Rib{\'o}}, {Rico},
  {Righi}, {Rugliancich}, {Saha}, {Sahakyan}, {Saito}, {Sakurai}, {Satalecka},
  {Schmidt}, {Schweizer}, {Sitarek}, {{\v{S}}nidari{\'c}}, {Sobczynska},
  {Somero}, {Stamerra}, {Strom}, {Strzys}, {Suda}, {Suri{\'c}}, {Takahashi},
  {Tavecchio}, {Temnikov}, {Terzi{\'c}}, {Teshima}, {Torres-Alb{\`a}}, {Tosti},
  {Tsujimoto}, {Vagelli}, {van Scherpenberg}, {Vanzo}, {Vazquez Acosta},
  {Vigorito}, {Vitale}, {Vovk}, {Will}, {Zari{\'c}}, \&
  {Nava}}]{2019Natur.575..455M}
{MAGIC Collaboration}, {Acciari}, V.~A., {Ansoldi}, S., {et~al.} 2019, \nat,
  575, 455, \dodoi{10.1038/s41586-019-1750-x}

\bibitem[{{Markey} \& {Tayler}(1973)}]{1973MNRAS.163...77M}
{Markey}, P., \& {Tayler}, R.~J. 1973, \mnras, 163, 77,
  \dodoi{10.1093/mnras/163.1.77}

\bibitem[{{Maselli} {et~al.}(2014){Maselli}, {Melandri}, {Nava}, {Mundell},
  {Kawai}, {Campana}, {Covino}, {Cummings}, {Cusumano}, {Evans}, {Ghirlanda},
  {Ghisellini}, {Guidorzi}, {Kobayashi}, {Kuin}, {La Parola}, {Mangano},
  {Oates}, {Sakamoto}, {Serino}, {Virgili}, {Zhang}, {Barthelmy}, {Beardmore},
  {Bernardini}, {Bersier}, {Burrows}, {Calderone}, {Capalbi}, {Chiang},
  {D'Avanzo}, {D'Elia}, {De Pasquale}, {Fugazza}, {Gehrels}, {Gomboc},
  {Harrison}, {Hanayama}, {Japelj}, {Kennea}, {Kopac}, {Kouveliotou}, {Kuroda},
  {Levan}, {Malesani}, {Marshall}, {Nousek}, {O'Brien}, {Osborne}, {Pagani},
  {Page}, {Page}, {Perri}, {Pritchard}, {Romano}, {Saito}, {Sbarufatti},
  {Salvaterra}, {Steele}, {Tanvir}, {Vianello}, {Weigand}, {Wiersema}, {Yatsu},
  {Yoshii}, \& {Tagliaferri}}]{2014Sci...343...48M}
{Maselli}, A., {Melandri}, A., {Nava}, L., {et~al.} 2014, Science, 343, 48,
  \dodoi{10.1126/science.1242279}

\bibitem[{{Melandri} {et~al.}(2019){Melandri}, {Izzo}, {D'Avanzo}, {Malesani},
  {Della Valle}, {Pian}, {Tanvir}, {Ragosta}, {Olivares}, {Carini}, {Palazzi},
  {Piranomonte}, {Jonker}, {Rossi}, {Kann}, {Hartmann}, {Inserra}, {Kankare},
  {Maguire}, {Smartt}, {Yaron}, {Young}, \& {Manulis}}]{2019GCN.23983....1G}
{Melandri}, A., {Izzo}, L., {D'Avanzo}, P., {et~al.} 2019, GRB Coordinates
  Network, 23983

\bibitem[{{Mestel}(1984)}]{1984AN....305..301M}
{Mestel}, L. 1984, Astronomische Nachrichten, 305, 301,
  \dodoi{10.1002/asna.2113050606}

\bibitem[{M{\'e}sz{\'a}ros \& Rees(1997)}]{1538-4357-482-1-L29}
M{\'e}sz{\'a}ros, P., \& Rees, M.~J. 1997, The Astrophysical Journal Letters,
  482, L29

\bibitem[{{Metzger} {et~al.}(1997){Metzger}, {Djorgovski}, {Kulkarni},
  {Steidel}, {Adelberger}, {Frail}, {Costa}, \&
  {Frontera}}]{1997Natur.387..878M}
{Metzger}, M.~R., {Djorgovski}, S.~G., {Kulkarni}, S.~R., {et~al.} 1997, \nat,
  387, 878, \dodoi{10.1038/43132}

\bibitem[{{Mirzoyan} {et~al.}(2019){Mirzoyan}, {Noda}, {Moretti}, {Berti},
  {Nigro}, {Hoang}, {Micanovic}, {Takahashi}, {Chai}, \&
  {Moralejo}}]{2019GCN.23701....1M}
{Mirzoyan}, R., {Noda}, K., {Moretti}, E., {et~al.} 2019, GRB Coordinates
  Network, 23701

\bibitem[{{Most} {et~al.}(2018){Most}, {Nathanail}, \&
  {Rezzolla}}]{2018ApJ...864..117M}
{Most}, E.~R., {Nathanail}, A., \& {Rezzolla}, L. 2018, \apj, 864, 117,
  \dodoi{10.3847/1538-4357/aad6ef}

\bibitem[{{Narayan} {et~al.}(1992){Narayan}, {Paczynski}, \&
  {Piran}}]{1992ApJ...395L..83N}
{Narayan}, R., {Paczynski}, B., \& {Piran}, T. 1992, \apjl, 395, L83,
  \dodoi{10.1086/186493}

\bibitem[{{Nathanail} {et~al.}(2017){Nathanail}, {Most}, \&
  {Rezzolla}}]{2017MNRAS.469L..31N}
{Nathanail}, A., {Most}, E.~R., \& {Rezzolla}, L. 2017, \mnras, 469, L31,
  \dodoi{10.1093/mnrasl/slx035}

\bibitem[{{Nomoto} {et~al.}(1994){Nomoto}, {Yamaoka}, {Pols}, {van den Heuvel},
  {Iwamoto}, {Kumagai}, \& {Shigeyama}}]{1994Natur.371..227N}
{Nomoto}, K., {Yamaoka}, H., {Pols}, O.~R., {et~al.} 1994, \nat, 371, 227,
  \dodoi{10.1038/371227a0}

\bibitem[{{Paczynski}(1991)}]{1991AcA....41..257P}
{Paczynski}, B. 1991, \actaa, 41, 257

\bibitem[{{Paczynski}(1992)}]{1992grbo.book...67P}
---. 1992, {Gamma-ray bursts from colliding neutron stars.}, ed. C.~{Ho}, R.~I.
  {Epstein}, \& E.~E. {Fenimore}, 67--74

\bibitem[{{Pian} {et~al.}(2006){Pian}, {Mazzali}, {Masetti}, {Ferrero},
  {Klose}, {Palazzi}, {Ramirez-Ruiz}, {Woosley}, {Kouveliotou}, {Deng},
  {Filippenko}, {Foley}, {Fynbo}, {Kann}, {Li}, {Hjorth}, {Nomoto}, {Patat},
  {Sauer}, {Sollerman}, {Vreeswijk}, {Guenther}, {Levan}, {O'Brien}, {Tanvir},
  {Wijers}, {Dumas}, {Hainaut}, {Wong}, {Baade}, {Wang}, {Amati}, {Cappellaro},
  {Castro-Tirado}, {Ellison}, {Frontera}, {Fruchter}, {Greiner}, {Kawabata},
  {Ledoux}, {Maeda}, {M{\o}ller}, {Nicastro}, {Rol}, \&
  {Starling}}]{2006Natur.442.1011P}
{Pian}, E., {Mazzali}, P.~A., {Masetti}, N., {et~al.} 2006, \nat, 442, 1011,
  \dodoi{10.1038/nature05082}

\bibitem[{{Pitts} \& {Tayler}(1985)}]{1985MNRAS.216..139P}
{Pitts}, E., \& {Tayler}, R.~J. 1985, \mnras, 216, 139,
  \dodoi{10.1093/mnras/216.2.139}

\bibitem[{{Price}(2011)}]{2011ascl.soft03004P}
{Price}, D.~J. 2011, {SPLASH: An Interactive Visualization Tool for Smoothed
  Particle Hydrodynamics Simulations}, Astrophysics Source Code Library.
\newblock \doeprint{1103.004}

\bibitem[{{Rees} \& {Meszaros}(1992)}]{1992MNRAS.258P..41R}
{Rees}, M.~J., \& {Meszaros}, P. 1992, \mnras, 258, 41P,
  \dodoi{10.1093/mnras/258.1.41P}

\bibitem[{{Rezzolla} {et~al.}(2011){Rezzolla}, {Giacomazzo}, {Baiotti},
  {Granot}, {Kouveliotou}, \& {Aloy}}]{2011ApJ...732L...6R}
{Rezzolla}, L., {Giacomazzo}, B., {Baiotti}, L., {et~al.} 2011, \apjl, 732, L6,
  \dodoi{10.1088/2041-8205/732/1/L6}

\bibitem[{{Riahi} {et~al.}(2019){Riahi}, {Kalantari}, \&
  {Rueda}}]{2019PhRvD..99d3004R}
{Riahi}, R., {Kalantari}, S.~Z., \& {Rueda}, J.~A. 2019, \prd, 99, 043004,
  \dodoi{10.1103/PhysRevD.99.043004}

\bibitem[{{Rowlinson} {et~al.}(2013){Rowlinson}, {O'Brien}, {Metzger},
  {Tanvir}, \& {Levan}}]{2013MNRAS.430.1061R}
{Rowlinson}, A., {O'Brien}, P.~T., {Metzger}, B.~D., {Tanvir}, N.~R., \&
  {Levan}, A.~J. 2013, \mnras, 430, 1061, \dodoi{10.1093/mnras/sts683}

\bibitem[{{Rowlinson} {et~al.}(2010){Rowlinson}, {O'Brien}, {Tanvir}, {Zhang},
  {Evans}, {Lyons}, {Levan}, {Willingale}, {Page}, {Onal}, {Burrows},
  {Beardmore}, {Ukwatta}, {Berger}, {Hjorth}, {Fruchter}, {Tunnicliffe}, {Fox},
  \& {Cucchiara}}]{2010MNRAS.409..531R}
{Rowlinson}, A., {O'Brien}, P.~T., {Tanvir}, N.~R., {et~al.} 2010, \mnras, 409,
  531, \dodoi{10.1111/j.1365-2966.2010.17354.x}

\bibitem[{{Rueda} \& {Ruffini}(2012)}]{2012ApJ...758L...7R}
{Rueda}, J.~A., \& {Ruffini}, R. 2012, \apjl, 758, L7,
  \dodoi{10.1088/2041-8205/758/1/L7}

\bibitem[{{Rueda} {et~al.}(2019){Rueda}, {Ruffini}, \&
  {Wang}}]{2019Univ....5..110R}
{Rueda}, J.~A., {Ruffini}, R., \& {Wang}, Y. 2019, Universe, 5, 110,
  \dodoi{10.3390/universe5050110}

\bibitem[{{Ruffini} {et~al.}(2018{\natexlab{a}}){Ruffini}, {Karlica},
  {Sahakyan}, {Rueda}, {Wang}, {Mathews}, {Bianco}, \&
  {Muccino}}]{2018ApJ...869..101R}
{Ruffini}, R., {Karlica}, M., {Sahakyan}, N., {et~al.} 2018{\natexlab{a}},
  \apj, 869, 101, \dodoi{10.3847/1538-4357/aaeac8}

\bibitem[{{Ruffini} {et~al.}(2019{\natexlab{a}}){Ruffini}, {Melon Fuksman}, \&
  {Vereshchagin}}]{2019ApJ...883..191R}
{Ruffini}, R., {Melon Fuksman}, J.~D., \& {Vereshchagin}, G.~V.
  2019{\natexlab{a}}, \apj, 883, 191, \dodoi{10.3847/1538-4357/ab3c51}

\bibitem[{Ruffini \& Wilson(1975)}]{Ruffini:1975ne}
Ruffini, R., \& Wilson, J.~R. 1975, Phys. Rev., D12, 2959,
  \dodoi{10.1103/PhysRevD.12.2959}

\bibitem[{{Ruffini} {et~al.}(2013){Ruffini}, {Bianco}, {Enderli}, {Muccino},
  {Penacchioni}, {Pisani}, {Rueda}, {Sahakyan}, {Wang}, \&
  {Izzo}}]{2013GCN..14526...1R}
{Ruffini}, R., {Bianco}, C.~L., {Enderli}, M., {et~al.} 2013, GCN Circ., 14526

\bibitem[{{Ruffini} {et~al.}(2015){Ruffini}, {Wang}, {Enderli}, {Muccino},
  {Kovacevic}, {Bianco}, {Penacchioni}, {Pisani}, \&
  {Rueda}}]{2015ApJ...798...10R}
{Ruffini}, R., {Wang}, Y., {Enderli}, M., {et~al.} 2015, \apj, 798, 10,
  \dodoi{10.1088/0004-637X/798/1/10}

\bibitem[{{Ruffini} {et~al.}(2016){Ruffini}, {Rueda}, {Muccino}, {Aimuratov},
  {Becerra}, {Bianco}, {Kovacevic}, {Moradi}, {Oliveira}, {Pisani}, \&
  {Wang}}]{2016ApJ...832..136R}
{Ruffini}, R., {Rueda}, J.~A., {Muccino}, M., {et~al.} 2016, \apj, 832, 136,
  \dodoi{10.3847/0004-637X/832/2/136}

\bibitem[{{Ruffini} {et~al.}(2018{\natexlab{b}}){Ruffini}, {Wang}, {Aimuratov},
  {Barres de Almeida}, {Becerra}, {Bianco}, {Chen}, {Karlica}, {Kovacevic},
  {Li}, {Melon Fuksman}, {Moradi}, {Muccino}, {Penacchioni}, {Pisani},
  {Primorac}, {Rueda}, {Shakeri}, {Vereshchagin}, \&
  {Xue}}]{2018ApJ...852...53R}
{Ruffini}, R., {Wang}, Y., {Aimuratov}, Y., {et~al.} 2018{\natexlab{b}}, \apj,
  852, 53, \dodoi{10.3847/1538-4357/aa9e8b}

\bibitem[{{Ruffini} {et~al.}(2018{\natexlab{c}}){Ruffini}, {Becerra}, {Bianco},
  {Chen}, {Karlica}, {Kova{\v c}evi{\'c}}, {Melon Fuksman}, {Moradi},
  {Muccino}, {Pisani}, {Primorac}, {Rueda}, {Vereshchagin}, {Wang}, \&
  {Xue}}]{2018ApJ...869..151R}
{Ruffini}, R., {Becerra}, L., {Bianco}, C.~L., {et~al.} 2018{\natexlab{c}},
  \apj, 869, 151, \dodoi{10.3847/1538-4357/aaee68}

\bibitem[{{Ruffini} {et~al.}(2018{\natexlab{d}}){Ruffini}, {Rodriguez},
  {Muccino}, {Rueda}, {Aimuratov}, {Barres de Almeida}, {Becerra}, {Bianco},
  {Cherubini}, {Filippi}, {Gizzi}, {Kovacevic}, {Moradi}, {Oliveira}, {Pisani},
  \& {Wang}}]{2018ApJ...859...30R}
{Ruffini}, R., {Rodriguez}, J., {Muccino}, M., {et~al.} 2018{\natexlab{d}},
  \apj, 859, 30, \dodoi{10.3847/1538-4357/aabee4}

\bibitem[{{Ruffini} {et~al.}(2018{\natexlab{e}}){Ruffini}, {Rueda}, {Moradi},
  {Wang}, {Xue}, {Becerra}, {Bianco}, {Chen}, {Cherubini}, {Filippi},
  {Karlica}, {Melon Fuksman}, {Primorac}, {Sahakyan}, \&
  {Vereshchagin}}]{2018arXiv181101839R}
{Ruffini}, R., {Rueda}, J.~A., {Moradi}, R., {et~al.} 2018{\natexlab{e}}, arXiv
  e-prints.
\newblock \doarXiv{1811.01839}

\bibitem[{{Ruffini} {et~al.}(2019{\natexlab{b}}){Ruffini}, {Li}, {Moradi},
  {Rueda}, {Wang}, {Xue}, {Bianco}, {Campion}, {Melon Fuksman}, {Cherubini},
  {Filippi}, {Karlica}, \& {Sahakyan}}]{2019arXiv190404162R}
{Ruffini}, R., {Li}, L., {Moradi}, R., {et~al.} 2019{\natexlab{b}}, arXiv
  e-prints.
\newblock \doarXiv{1904.04162}

\bibitem[{{Ruffini} {et~al.}(2019{\natexlab{c}}){Ruffini}, {Moradi}, {Rueda},
  {Becerra}, {Bianco}, {Cherubini}, {Filippi}, {Chen}, {Karlica}, {Sahakyan},
  {Wang}, \& {Xue}}]{2019ApJ...886...82R}
{Ruffini}, R., {Moradi}, R., {Rueda}, J.~A., {et~al.} 2019{\natexlab{c}}, \apj,
  886, 82, \dodoi{10.3847/1538-4357/ab4ce6}

\bibitem[{{Ruffini} {et~al.}(2019{\natexlab{d}}){Ruffini}, {Moradi},
  {Aimuratov}, {Barres de Almeida}, {Belinski}, {Bianco}, {Chen}, {Cherubini},
  {Filippi}, {Fuksman}, {Karlica}, {Li}, {Primorac}, {Rueda}, {Sakahyan},
  {Wang}, \& {Xue}}]{2019GCN.23715....1R}
{Ruffini}, R., {Moradi}, R., {Aimuratov}, Y., {et~al.} 2019{\natexlab{d}}, GRB
  Coordinates Network, 23715

\bibitem[{{Sari}(1997)}]{1997ApJ...489L..37S}
{Sari}, R. 1997, \apjl, 489, L37, \dodoi{10.1086/310957}

\bibitem[{{Sari} \& {Piran}(1995)}]{1995ApJ...455L.143S}
{Sari}, R., \& {Piran}, T. 1995, \apjl, 455, L143, \dodoi{10.1086/309835}

\bibitem[{{Sari} {et~al.}(1998){Sari}, {Piran}, \&
  {Narayan}}]{1998ApJ...497L..17S}
{Sari}, R., {Piran}, T., \& {Narayan}, R. 1998, \apjl, 497, L17,
  \dodoi{10.1086/311269}

\bibitem[{{Selsing} {et~al.}(2019){Selsing}, {Fynbo}, {Heintz}, {Watson}, \&
  {Dyrbye}}]{2019GCN.23695....1S}
{Selsing}, J., {Fynbo}, J.~P.~U., {Heintz}, K.~E., {Watson}, D., \& {Dyrbye},
  N. 2019, GRB Coordinates Network, 23695

\bibitem[{{Shibata} {et~al.}(2006){Shibata}, {Duez}, {Liu}, {Shapiro}, \&
  {Stephens}}]{2006PhRvL..96c1102S}
{Shibata}, M., {Duez}, M.~D., {Liu}, Y.~T., {Shapiro}, S.~L., \& {Stephens},
  B.~C. 2006, Physical Review Letters, 96, 031102,
  \dodoi{10.1103/PhysRevLett.96.031102}

\bibitem[{Skilling(2004)}]{doi:10.1063/1.1835238}
Skilling, J. 2004, AIP Conference Proceedings, 735, 395,
  \dodoi{10.1063/1.1835238}

\bibitem[{{Spruit}(1999)}]{1999A&A...349..189S}
{Spruit}, H.~C. 1999, \aap, 349, 189

\bibitem[{{Spruit}(2009)}]{2009IAUS..259...61S}
{Spruit}, H.~C. 2009, in IAU Symposium, Vol. 259, Cosmic Magnetic Fields: From
  Planets, to Stars and Galaxies, ed. K.~G. {Strassmeier}, A.~G. {Kosovichev},
  \& J.~E. {Beckman}, 61--74, \dodoi{10.1017/S1743921309030075}

\bibitem[{{Stephens} {et~al.}(2007){Stephens}, {Duez}, {Liu}, {Shapiro}, \&
  {Shibata}}]{2007CQGra..24S.207S}
{Stephens}, B.~C., {Duez}, M.~D., {Liu}, Y.~T., {Shapiro}, S.~L., \& {Shibata},
  M. 2007, Classical and Quantum Gravity, 24, S207,
  \dodoi{10.1088/0264-9381/24/12/S14}

\bibitem[{{Stephens} {et~al.}(2008){Stephens}, {Shapiro}, \&
  {Liu}}]{2008PhRvD..77d4001S}
{Stephens}, B.~C., {Shapiro}, S.~L., \& {Liu}, Y.~T. 2008, \prd, 77, 044001,
  \dodoi{10.1103/PhysRevD.77.044001}

\bibitem[{{Sturner} {et~al.}(1997){Sturner}, {Skibo}, {Dermer}, \&
  {Mattox}}]{1997ApJ...490..619S}
{Sturner}, S.~J., {Skibo}, J.~G., {Dermer}, C.~D., \& {Mattox}, J.~R. 1997,
  \apj, 490, 619, \dodoi{10.1086/304894}

\bibitem[{{Tam} {et~al.}(2017){Tam}, {He}, {Tang}, \&
  {Wang}}]{2017ApJ...844L...7T}
{Tam}, P.-H.~T., {He}, X.-B., {Tang}, Q.-W., \& {Wang}, X.-Y. 2017, \apjl, 844,
  L7, \dodoi{10.3847/2041-8213/aa7ca5}

\bibitem[{{Tanvir} {et~al.}(2016){Tanvir}, {Levan}, {Cenko}, {Perley},
  {Cucchiara}, {Roth}, {Wiersema}, {Fruchter}, \&
  {Laskar}}]{2016GCN.19419....1T}
{Tanvir}, N.~R., {Levan}, A.~J., {Cenko}, S.~B., {et~al.} 2016, GRB Coordinates
  Network, Circular Service, No.~19419, \#1 (2016), 19419

\bibitem[{{Tayler}(1973)}]{1973MNRAS.161..365T}
{Tayler}, R.~J. 1973, \mnras, 161, 365, \dodoi{10.1093/mnras/161.4.365}

\bibitem[{{Tayler}(1980)}]{1980MNRAS.191..151T}
---. 1980, \mnras, 191, 151, \dodoi{10.1093/mnras/191.1.151}

\bibitem[{{Troja} {et~al.}(2007){Troja}, {Cusumano}, {O'Brien}, {Zhang},
  {Sbarufatti}, {Mangano}, {Willingale}, {Chincarini}, {Osborne}, {Marshall},
  {Burrows}, {Campana}, {Gehrels}, {Guidorzi}, {Krimm}, {La Parola}, {Liang},
  {Mineo}, {Moretti}, {Page}, {Romano}, {Tagliaferri}, {Zhang}, {Page}, \&
  {Schady}}]{2007ApJ...665..599T}
{Troja}, E., {Cusumano}, G., {O'Brien}, P.~T., {et~al.} 2007, \apj, 665, 599,
  \dodoi{10.1086/519450}

\bibitem[{{Vianello} {et~al.}(2017){Vianello}, {Lauer}, {Burgess}, {Ayala},
  {Fleischhack}, {Harding}, {Hui}, {Marinelli}, {Savchenko}, \&
  {Zhou}}]{2017ifs..confE.130V}
{Vianello}, G., {Lauer}, R.~J., {Burgess}, J.~M., {et~al.} 2017, in Proceedings
  of the 7th International Fermi Symposium, 130

\bibitem[{{Wald}(1974)}]{1974PhRvD..10.1680W}
{Wald}, R.~M. 1974, \prd, 10, 1680, \dodoi{10.1103/PhysRevD.10.1680}

\bibitem[{{Wang} {et~al.}(2019{\natexlab{a}}){Wang}, {Li}, {Moradi}, \&
  {Ruffini}}]{2019arXiv190107505W}
{Wang}, Y., {Li}, L., {Moradi}, R., \& {Ruffini}, R. 2019{\natexlab{a}}, arXiv
  e-prints.
\newblock \doarXiv{1901.07505}

\bibitem[{{Wang} {et~al.}(2019{\natexlab{b}}){Wang}, {Rueda}, {Ruffini},
  {Becerra}, {Bianco}, {Becerra}, {Li}, \& {Karlica}}]{2019ApJ...874...39W}
{Wang}, Y., {Rueda}, J.~A., {Ruffini}, R., {et~al.} 2019{\natexlab{b}}, \apj,
  874, 39, \dodoi{10.3847/1538-4357/ab04f8}

\bibitem[{{Waxman} \& {Piran}(1994)}]{1994ApJ...433L..85W}
{Waxman}, E., \& {Piran}, T. 1994, \apjl, 433, L85, \dodoi{10.1086/187554}

\bibitem[{{Wijers} {et~al.}(1997){Wijers}, {Rees}, \&
  {Meszaros}}]{1997MNRAS.288L..51W}
{Wijers}, R. A.~M.~J., {Rees}, M.~J., \& {Meszaros}, P. 1997, \mnras, 288, L51,
  \dodoi{10.1093/mnras/288.4.L51}

\bibitem[{{Wilson}(1975)}]{1975NYASA.262..123W}
{Wilson}, J.~R. 1975, in Annals of the New York Academy of Sciences, Vol. 262,
  Seventh Texas Symposium on Relativistic Astrophysics, ed. P.~G. {Bergman},
  E.~J. {Fenyves}, \& L.~{Motz}, 123--132,
  \dodoi{10.1111/j.1749-6632.1975.tb31425.x}

\bibitem[{{Wilson}(1978)}]{1978pans.proc..644W}
{Wilson}, J.~R. 1978, in Physics and Astrophysics of Neutron Stars and Black
  Holes, ed. R.~{Giacconi} \& R.~{Ruffini}, 644--675

\bibitem[{{Woosley}(1993)}]{1993ApJ...405..273W}
{Woosley}, S.~E. 1993, \apj, 405, 273, \dodoi{10.1086/172359}

\bibitem[{{Woosley} \& {Bloom}(2006)}]{2006ARA&A..44..507W}
{Woosley}, S.~E., \& {Bloom}, J.~S. 2006, \araa, 44, 507,
  \dodoi{10.1146/annurev.astro.43.072103.150558}

\bibitem[{{Wright}(1973)}]{1973MNRAS.162..339W}
{Wright}, G.~A.~E. 1973, \mnras, 162, 339, \dodoi{10.1093/mnras/162.4.339}

\bibitem[{{Xu} {et~al.}(2016){Xu}, {Malesani}, {Fynbo}, {Tanvir}, {Levan}, \&
  {Perley}}]{2016GCN.19600....1X}
{Xu}, D., {Malesani}, D., {Fynbo}, J.~P.~U., {et~al.} 2016, GRB Coordinates
  Network, Circular Service, No.~19600, \#1 (2016), 19600

\bibitem[{{Xu} {et~al.}(2013){Xu}, {de Ugarte Postigo}, {Leloudas},
  {Kr{\"u}hler}, {Cano}, {Hjorth}, {Malesani}, {Fynbo}, {Th{\"o}ne},
  {S{\'a}nchez-Ram{\'{\i}}rez}, {Schulze}, {Jakobsson}, {Kaper}, {Sollerman},
  {Watson}, {Cabrera-Lavers}, {Cao}, {Covino}, {Flores}, {Geier}, {Gorosabel},
  {Hu}, {Milvang-Jensen}, {Sparre}, {Xin}, {Zhang}, {Zheng}, \&
  {Zou}}]{2013ApJ...776...98X}
{Xu}, D., {de Ugarte Postigo}, A., {Leloudas}, G., {et~al.} 2013, \apj, 776,
  98, \dodoi{10.1088/0004-637X/776/2/98}

\bibitem[{{Yoshida} \& {Umeda}(2011)}]{2011MNRAS.412L..78Y}
{Yoshida}, T., \& {Umeda}, H. 2011, \mnras, 412, L78,
  \dodoi{10.1111/j.1745-3933.2011.01008.x}

\bibitem[{{Zhang}(2018)}]{2018pgrb.book.....Z}
{Zhang}, B. 2018, {The Physics of Gamma-Ray Bursts},
  \dodoi{10.1017/9781139226530}

\bibitem[{{Zhang} \& {M{\'e}sz{\'a}ros}(2001)}]{2001ApJ...552L..35Z}
{Zhang}, B., \& {M{\'e}sz{\'a}ros}, P. 2001, \apjl, 552, L35,
  \dodoi{10.1086/320255}

\bibitem[{{Zhang} {et~al.}(2018){Zhang}, {Zhang}, {Castro-Tirado}, {Dai},
  {Tam}, {Wang}, {Hu}, {Karpov}, {Pozanenko}, {Zhang}, {Mazaeva}, {Minaev},
  {Volnova}, {Oates}, {Gao}, {Wu}, {Shao}, {Tang}, {Beskin}, {Biryukov},
  {Bondar}, {Ivanov}, {Katkova}, {Orekhova}, {Perkov}, {Sasyuk}, {Mankiewicz},
  {{\.Z}arnecki}, {Cwiek}, {Opiela}, {Zadro{\.Z}ny}, {Aptekar}, {Frederiks},
  {Svinkin}, {Kusakin}, {Inasaridze}, {Burhonov}, {Rumyantsev}, {Klunko},
  {Moskvitin}, {Fatkhullin}, {Sokolov}, {Valeev}, {Jeong}, {Park},
  {Caballero-Garc{\'{\i}}a}, {Cunniffe}, {Tello}, {Ferrero}, {Pandey},
  {Jel{\'{\i}}nek}, {Peng}, {S{\'a}nchez-Ram{\'{\i}}rez}, \&
  {Castell{\'o}n}}]{2018NatAs...2...69Z}
{Zhang}, B.-B., {Zhang}, B., {Castro-Tirado}, A.~J., {et~al.} 2018, Nature
  Astronomy, 2, 69, \dodoi{10.1038/s41550-017-0309-8}

\end{thebibliography}

\end{document}